\documentclass[onecolumn]{IEEEtran}
\usepackage{cite}
\usepackage{amsmath,amssymb,amsfonts}
\usepackage{algorithmic}
\usepackage{textcomp}
\usepackage[utf8]{inputenc}
\usepackage[T1]{fontenc} 
\usepackage{mathtools}
\usepackage{graphicx}
\usepackage{tabularx}
\usepackage{subfigure}
\usepackage[table]{xcolor}
\graphicspath{{images/}}
\usepackage{dblfloatfix}
\usepackage{fixltx2e}
\usepackage{multirow}
\usepackage[justification=centering]{caption}
\usepackage{array}	
\usepackage{algorithm}
\usepackage{lineno,hyperref}
\usepackage{bytefield}
\usepackage{morefloats}
\newsavebox{\bytefieldbox}
\usepackage{comment}
\usepackage{moreverb}
\usepackage{hyperref}%
\usepackage{xparse}%
\usepackage{morewrites}
\usepackage{wrapfig}
\usepackage{xkeyval}%

\begin{document}

\title{Collaboratively assessing urban alerts in ad hoc participatory sensing}

\author{\IEEEauthorblockN{F\'atima Castro-Jul}
\IEEEauthorblockA{AtlantTTic Research Center, University of Vigo, Spain.
Email: avilas@det.uvigo.es}\\
\and
\IEEEauthorblockN{Rebeca Díaz Redondo}
\IEEEauthorblockA{AtlantTTic Research Center, University of Vigo, Spain.
Email: rebeca@det.uvigo.es}\\
\and
\IEEEauthorblockN{Ana Fern\'andez-Vilas}
\IEEEauthorblockA{AtlantTTic Research Center, University of Vigo, Spain.
Email: avilas@det.uvigo.es}}
\maketitle

\begin{abstract}
Ad hoc architectures have emerged as a valuable alternative to centralized participatory sensing systems due to their infrastructure-less nature, which ensures good availability, easy maintenance and direct user communication. As a result, they need to incorporate content-aware assessment mechanisms to deal with a common problem in participatory sensing: information assessment. Easy contribution encourages users' participation and improves the sensing task but may result in large amounts of data, which may not be valid or relevant. Currently, prioritization is the only totally ad hoc scheme to assess user-generated alerts. This strategy prevents duplicates from congesting the network. However, it does not include the assessment of every generated alert and does not deal with low-quality or irrelevant alerts.\\ 
In order to ensure users receive only interesting alerts and the network is not compromised, we propose two collaborative alert assessment mechanisms that, while keeping the network flat, provide an effective message filter. Both of them rely on opportunistic collaboration with nearby peers. By simulating their behavior in a real urban area, we have proved them able to decrease network load while maintaining alert delivery ratio. 
 
\end{abstract}


\section{Introduction}
The Smart City paradigm conceives the city as an intelligent and connected environment where information technologies are embedded everywhere, constantly monitoring what is going on in order inform about and take care of any incident that may take place. The rise of sensor-rich smartphones, ubiquitous nowadays, has turned citizens into sensors of their own environment \cite{Campbell2006}, enabling the birth of participatory sensing \cite{Burke2006} and crowdsensing schemes \cite{Guo2014, Guo2015}. Since these systems rely solely on users' contribution to gather context information, they do not require a dedicated infrastructure and therefore, they are an inexpensive alternative to collect data about the whole city. Users' involvement makes sensing systems flexible and rich as diverse views can be considered. Therefore, users' participation should be encouraged and simplified to maximize data gathering and its diversity. However, easy contribution may result in large amounts of information, which may be incorrect, irrelevant, low-quality \cite{Zhang2014} or redundant \cite{Uddin2011}.\\ 
Data quality assessment is a problem common to every grassroots-based scheme but it becomes more complex in systems where there is no central entity in charge of processing the gathered information. That is the case of totally ad hoc participatory sensing \cite{Uddin2011, Cacciapuoti2013}, where users communicate directly with each other in a distributed way. These systems are generally targeted at spreading event alerts to nearby users in highly-populated urban areas or when traditional networks are challenged. In absence of a filtering entity that decides what is worthy to be disseminated, every generated message is sent to all users, whether they are interested or not. As a consequence, multiple messages increase network load without adding any benefit. In a crowded scenario where many users are spreading news about a certain event, the network may get congested and collapse. As a result, some users may not receive the alerts. Ironically, too much information may prevent them from being informed. There exist solutions \cite{Sun2003} based on hierarchical mechanisms, which reduce the problem while imposing a setup cost that makes the system not as flexible. An alternative that maintains the network flat is content-aware prioritization \cite{Uddin2011}. However, this strategy focuses only in redundant information and does not target messages dissimilar to others but of low-quality. Moreover, it does not evaluate every message but it is only triggered when a significant number of messages is already in the network. As a result, it does not deal with the problem thoroughly.\\
To deal with the problem of low-quality data in ad hoc participatory sensing and prevent unnecessary transmissions, we have developed collaboration schemes that involve neighbor nodes in the assessment of information pertinence. Unlike other systems, assessment is performed in a distributed manner without any centralized entity or hierarchical structure. Moreover, it includes the evaluation of every user-generated message before being disseminated. Our objective is to reduce the network load required to inform other users on a certain event. We aim to do so by reducing duplicate and irrelevant messages; in other words, by improving the quality of the participatory sensed information. In this paper, we present the design and evaluation of two different collaborative mechanisms, which differ in the level of neighbor cooperation they require. One of them had already been briefly introduced in our previous work \cite{Castro-Jul2016}. We examine in depth the problems of information assessment and opportunistic collaboration in participatory sensing and we detail the operation of both strategies. Furthermore, we have conducted network simulations based on real urban scenarios in order to assess their performance.\\  
This paper is structured as follows. In Section \ref{sec:related_work}, we start by providing an overview of participatory alert systems and we discuss how information assessment and opportunistic collaboration have been dealt with in this field. Then, we describe our target scenario (Section \ref{sec:scenario}) and we detail our approach (Section \ref{sec:approach}). In Section \ref{sec:simulation} we describe the evaluation process, which is discussed in Section \ref{sec:evaluation}. Finally, we finish by outlining our conclusions and future work (Section \ref{sec:conclusion}).

\section{Participatory alert systems} \label{sec:related_work}
The aim of participatory and crowdsensing is to collectively gather information. This information is employed mainly for two purposes: \textit{decision making} and \textit{visualization and sharing} \cite{Guo2015}. The former approach is meant to form collaborative knowledge, to be used by the very contributors or by a central entity in charge of aggregating and processing the data. Decision making applications include object recognition, recommendation and prediction. In systems targeted at  \textit{visualization and sharing} the information gathered is meant to be distributed to other users besides the contributors. Applications include sharing users' monitoring results to motivate them by competing with peers \cite{Fujiki2008}, creating collaborative knowledge to be made publicly available \cite{DHondt2013} and disseminating information on detected events. These last systems are alert systems, particularly interesting in smart cities, where citizens benefit from being always aware of what is happening around them. In these systems, users disseminate alerts, which are notifications of events or incidents that draw their attention. The range of possible alert topic is broad, from earthquakes \cite{Sakaki2010} to potholes \cite{Eriksson2008}. Moreover, it includes not only potential unsafe situations but any other kind of unusual happening, like street shows or demonstrations. There exist a great variety of alert mechanisms, considering different architectures and degrees of user involvement \cite{Guo2015}. According to these two criteria, we establish the following categories of event detection and dissemination schemes: 

\begin{itemize}
\item \textbf{Centralized explicit}. Users with a specific application in their phones share information with a central entity in charge of gathering and processing the information. Information can be sent manually by users themselves or automatically when a certain sensor reading is triggered \cite{Eriksson2008}. Once processed, the information is made publicly available \cite{DHondt2013} or sent to users that have either subscribed to the service or queried for information \cite{Zhou2012}.
\item \textbf{Centralized implicit}. Users share information on events but not with the explicit intention of contributing to the system. In other words, they are not aware of the sensing task. Event detection is normally performed on online social networks information using data mining techniques to find meaningful events. As in the centralized explicit scheme, information can be spread to interested users \cite{Sakaki2010} or publicly distributed.
\item \textbf{Ad hoc}. Users detect events and share information directly with their physical neighbors, using short-range communication technologies \cite{Uddin2011,Cacciapuoti2013}. Thus, ad hoc systems are always explicit and do not include a central entity in charge of data gathering and processing. However, users may form hierarchical structures or clusters, with a leader that plays this role locally \cite{Sun2003}.
\end{itemize}

\begin{figure}
\centering
\begin{minipage}[c]{\linewidth}
\centering
\includegraphics[width=0.7\linewidth]{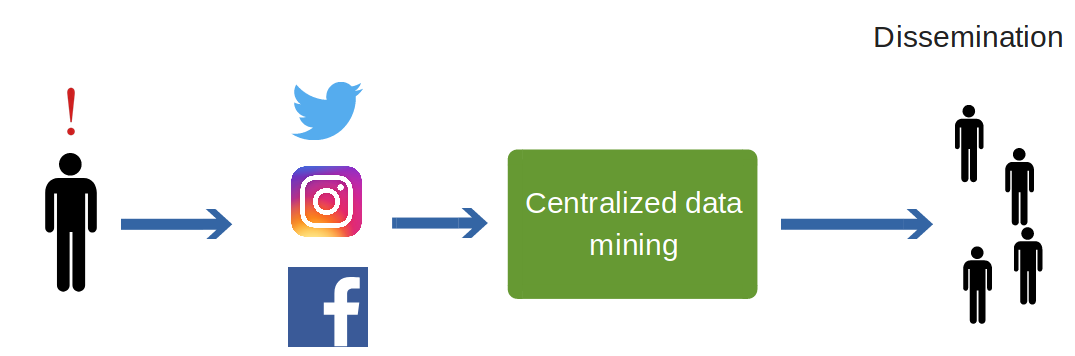}
\caption{Centralized implicit\label{subfig:centralized-implicit}}
\end{minipage}
\begin{minipage}[b]{.6\linewidth}
\centering
\includegraphics[width=1\textwidth]{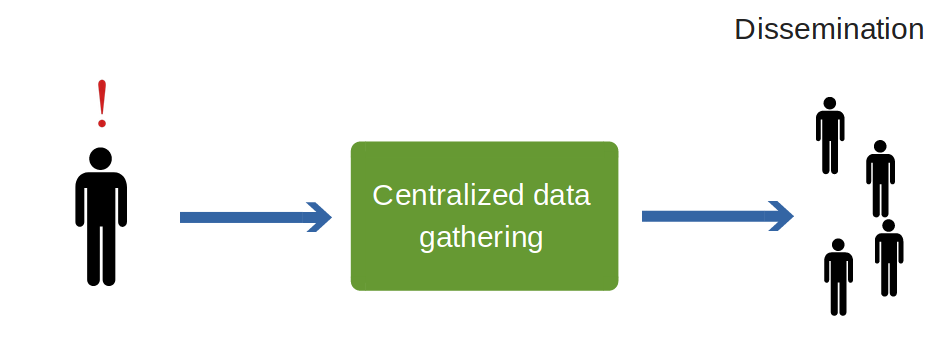}
\caption{Centralized explicit\label{subfig:centralized-explicit}}
\end{minipage}
\begin{minipage}[b]{.35\linewidth}
\centering
\includegraphics[width=1\textwidth]{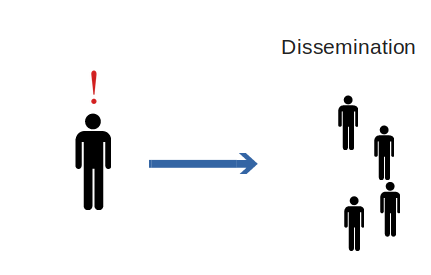}
\caption{Ad hoc\label{subfig:ad-hoc}}
\end{minipage}%
\caption{Participatory sensing alert systems\label{fig:alert-process}}
\end{figure}

What we define here as \textit{alert systems} has been referred to in the literature with different terms, such as situation awareness systems \cite{Uddin2011}. Systems based on social networks are described as event detection systems since they often do not deal with the dissemination phase. Explicit systems are commonly described simply as participatory sensing, without mention to the application they target. We consider all these systems as a whole since, even though they may not thoroughly cover the alert process, they play a part in leveraging user-generated information to detect events and spread the word about them. The architecture of a complete alert system using every one of the architectures above is shown in Figure \ref{fig:alert-process}.

\subsection{Information assessment in participatory sensing}
Event detection and dissemination have attracted researchers’ attention
over the last few years. Due to the big amount of data available nowadays, special attention has been paid to data mining in social networks \cite{Sakaki2010,Imran2015,Li2012a}. Data from social networks, especially Twitter, is processed in order to find trends or bursts of information \cite{Mathioudakis2010,Marcus2011} that may correspond to relevant events. Supervised classification is another usual detection technique \cite{Ritter2012}. Thus, these approaches benefit from handling as much data as possible. Irrelevant data is discarded and deduplication, performed using clustering methods, is only considered as a preprocessing method to decrease the amount of messages to be dealt with \cite{Imran2015,Rogstadius2013}.\\
Explicit alert systems rely also on a central entity in charge of processing gathered data and therefore of assessing alerts by comparing them with others. They may assume all them as correct and always trigger the alert, but only the first time it is received to prevent duplicates. Other processing techniques can include only triggering the alert if a certain number of users has corroborated the information \cite{Eriksson2008}. Anyway, this strategy also benefits from receiving as much data as possible since it improves event recognition and does not imply flooding the network.\\
In contrast, ad hoc alert systems do not benefit from large amounts of data as they may cause problems and even the collapse of the network. Research focus in this area is, in fact, targeted at how to decrease the amount of messages sent while maintaining the ratio of delivered information. In the alert case, this implies sending fewer alert messages while not missing relevant events. Work in this field has been mostly focused on efficient routing or dissemination techniques \cite{Bur2011,Goyal2012,Kokuti2012,Tonguz2006}, instead of filtering the content that is being disseminated. Messages are, in general, assumed to be relevant and it is up to the nodes to decide whether they should start alert propagation.\\
However, the idea of implementing filtering techniques based on message content is not new. The SPIN family of protocols \cite{Heinzelman1999,Kulik2002, Rehena2011} includes negotiation mechanisms where nodes receive first an update with a metadata description of available information so they can decide if they are interested in it or are able to receive them taking into account energy constraints. Instead of negotiating every delivery, interested-based dissemination can also be performed employing users' predefined interests \cite{Li2014}.
Another approach to message reduction are hierarchical protocols. Nodes are organized in clusters where the head is in charge of information transmission. It acts as a gateway to send data outside the local group, by its own judgment or answering to external requests. Node clustering can be based on physical neighborhood determined by the transmission technology or on previously defined geographical areas \cite{Sun2003,Saleet2007,Tsai2012}. In vehicle networks, gateway role is sometimes played by road side units (RSU) \cite{Tsai2012}. Data to be disseminated can be chosen by voting or directly by the leader, that may apply data fusion or aggregated techniques, such as correlation. \\
The only alternative that does not include a hierarchy or negotiation in every step is content-aware prioritization \cite{Uddin2011}. This strategy is applied on delay tolerant networks (DTN), where users carrying messages have a limited space for them and will, therefore, delete those most similar to others when storage space is scarce. Thus, it reduces the number of alerts by removing the redundant ones. However, it only evaluates messages when space constrains have been met and does not consider other kinds of low-quality data, such as incorrect or irrelevant notifications.

\subsection{Opportunistic collaboration in participatory sensing}
Opportunistic contacts are established between devices that are in the range of a short-range communication technology, normally Bluetooth or ad hoc WiFi, both of them widely present in smartphones. Thus, nearby devices can directly share data without a network infrastructure. Opportunistic collaboration exploits human mobility and proximity contacts by creating dynamic groups of co-located devices \cite{Conti2010a} with two aims in mind. On the one hand, nearby devices are the ones most likely to be interested in what is happening around them and therefore in receiving information on local issues \cite{Meier2002}. On the other hand, devices in proximity can obtain similar ambient information and therefore they are able to confirm or dismiss context information created by others in their vicinity. The former proximity application is targeted at sharing information in the vicinity itself, while the latter uses proximity to obtain collaborative data to be further forwarded, generally to a cloud entity. Applications that are limited to the vicinity include sharing information on traffic events \cite{Meier2002, Koukoumidis2011} and collaboratively deciding what is the appropriate device behavior in a certain event or situation \cite{Castro-Jul2017a}. \\
Collaboration in data gathering applications may be designed to obtain more accurate or enhanced information. To do so, users share their sensor readings, information about the models they employ for measurements \cite{Miluzzo2010} or extra data to augment the information obtained by others, for instance by helping them to tag the pictures they take in a certain event \cite{Qin2014}. Collaboration can also be targeted at reducing energy expenditure by allocating different sensing tasks or sampling times to different devices \cite{Lee2012, Shi2011}. Thus, the sensing task is neither duplicated nor falls on an only sensor. All the devices are assumed to be able to provide the same, valid information.\\
Ad hoc networks are an extension of opportunistic collaboration for larger areas. Beyond small groups formed at short-range distances, opportunistic collaboration focuses mostly on enabling communication between devices that are in the same densely-populated area and can be reached in several short-range hops. Traditionally, ad hoc networks have been targeted at emergency situations, where traditional communications are challenged, being employed as the only communication system or together with others \cite{Harkous2011}. However, they are also useful in others scenarios \cite{Conti2010a}, such as urban transportation systems and research conferences \cite{Li2014}.\\
The possibilities of short-range collaboration have never been exploited in a system targeted at optimizing the ad hoc dissemination of locally-inferred data.

\subsection{Opportunistic collaboration for information assessment}
We have identified the lack of comprehensive and totally ad hoc information assessment strategies. The absence of an adequate alert evaluation mechanism endangers the progress of ad hoc participatory sensing systems as a result of the distrust of user-generated unfiltered content and the risk of network collapse due to information abundance. As a result, it is imperative to develop assessment techniques for user-generated local information. Collaboration with users at a short-range distance has a great potential for this task, unexploited so far.\\
Our work is based on the fact that nearby users are able to access the same ambient data. Therefore, they are aware of the same circumstances, can detect the same incidents and are able to judge the appropriateness of their dissemination for the purpose of the system. As a consequence, if an incident alert cannot be confirmed by users in the short-range vicinity of the alert sender, it can be concluded that the alert should not be disseminated. \\

\begin{figure}
\centering
\includegraphics[width=1\textwidth]{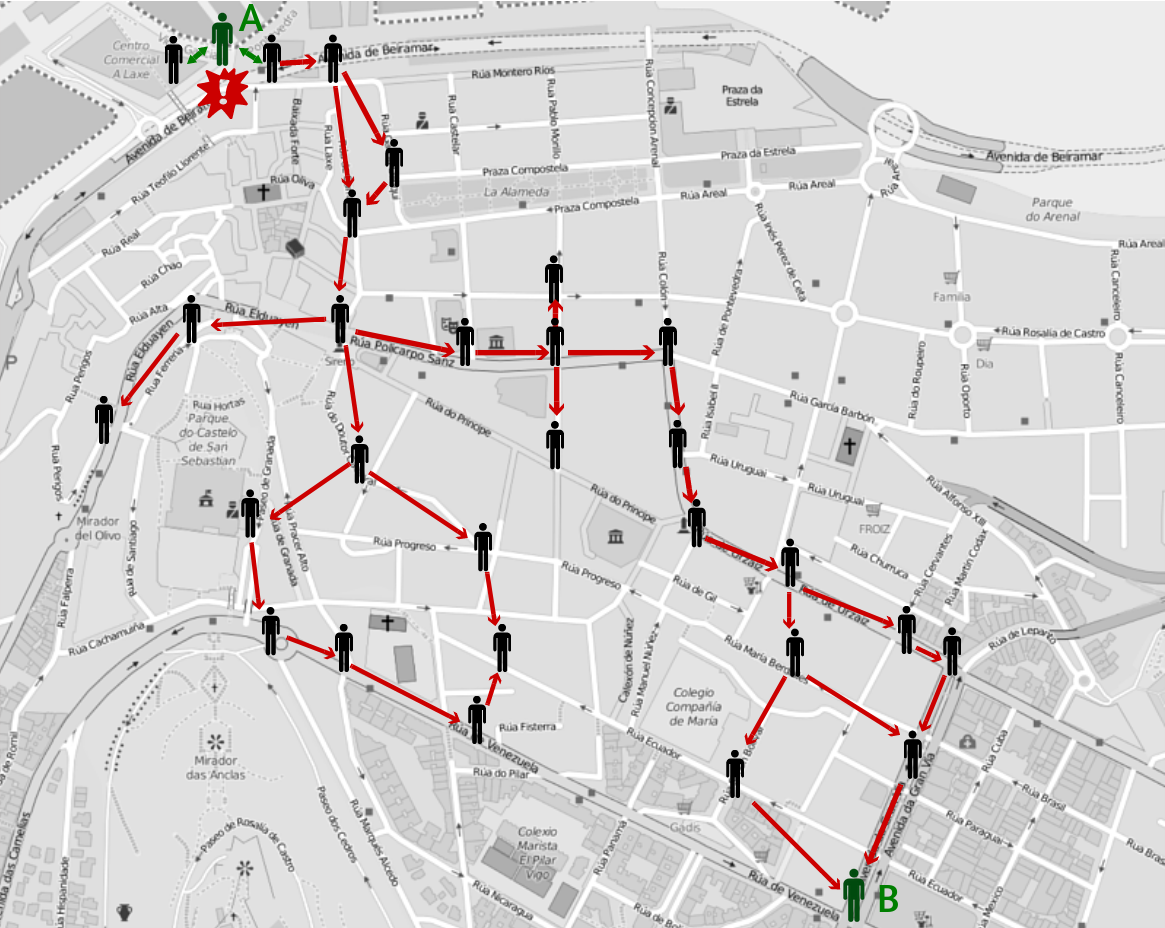}
\caption{Urban area used in simulation. Dissemination strategy and nodes considered to measure transmission time to opposite edge are shown.\label{map}}
\end{figure}

\begin{figure}
\centering
\includegraphics[width=0.9\linewidth]{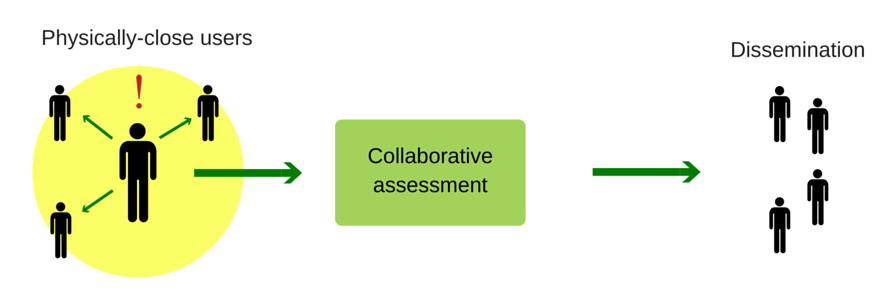}
\caption{Ad hoc alert system with collaborative assessment\label{fig:collaborative-assessment}}
\end{figure}

\section{Scenario} \label{sec:scenario}
Being aware of what is going on in the area allows citizens to best adapt to the changing circumstances of the city. For instance, they can modify their path to avoid a road that has been closed due to a traffic accident or to join a street show that fits their interests.\\
As a case of study, we have chosen an urban scenario with a high density of pedestrians that walk through the city employing a mobile application that designs the best route for them anytime. In order to plan the itinerary, the application takes into account alerts received from other users that report incidents in their close surroundings. Citizens act both as alert senders and recipients. Thus, they both contribute to the alert system and employ the navigation system enhanced with collaborative information. \\
Users' handheld devices are equipped with different sensors that allow them to monitor their surroundings and with short-range communication technologies, such as Bluetooth and ad hoc WiFi, that enable them to distribute their findings directly with their peers. Our alert system is targeted at general interest alerts, such as traffic incidents, public transport issues, crowds or unsafe situations, such as robberies or fights. Thus, we ensure users' concern and ability to discern them. The alert process may be triggered manually by users or automatically by devices. For instance, a car incident or a big crowd can be detected using smartphone microphones, which have proved to successfully act as sound sensors \cite{DHondt2013}. Since devices can communicate directly with each other, no dedicated communicated infrastructure is required. Connections build a flat network, where no hierarchies or clusters are formed since the high pedestrian mobility does not make setup costs worth the effort. The only network mechanism implemented is a simple hello mechanism for one-hop neighbor count. Connections are flexible and proximity-based, which makes it simple for any device to join and contribute to the detection and dissemination process.\\ 
We have targeted our contribution at an explicit ad hoc participatory sensing system for a dynamic urban scenario with a high density of pedestrians carrying mobile devices. This scenario enables direct ad hoc dissemination but the network is vulnerable to be flooded with unnecessary messages if no assessment mechanism is included.

\section{Collaboratively-triggered dissemination} \label{sec:approach}
We propose a collaboratively-triggered dissemination (CTD) mechanism composed of three steps: user alert, collaborative assessment and alert dissemination. When an incident is detected, a broadcast notification is sent to neighbor nodes in order to be assessed and, if approved, further disseminated.\\
Our contribution focuses on how to assess alert relevance and prevent alert duplication by using ad-hoc device cooperation. The evaluation mechanisms we have designed are detailed in Section \ref{subsec:proposal}. Both of them are based on proximity-based communications, restricting collaboration to physically close devices. Thus, only nodes with actual knowledge of the area, and therefore able to confirm or refute the alert, take part in the assessment process. For this reason, there is no need for security mechanisms to prevent intruders from interfering. This represents an advantage regarding social network approaches, where it is necessary to verify alert senders' location. Alert duplication is prevented by requiring devices to remember the latest alerts they have received. In this manner, they will not back up the same alert twice or start a new dissemination if they detect the same event. However, if an alert has been received through flooding dissemination, instead of a direct request for assessment, the node will start its own assessment process as the alert has been originated outside its local area. \\
If the alert is backed up by neighbors, it is disseminated through the whole network using controlled flooding. Since we aim our attention to the assessment stage, developing an efficient strategy in the dissemination phase is out of the scope of this paper. However, our assessment strategy could be included before any other dissemination technique such as a DTN. \\
Alert verification takes part exclusively in the first sender's neighborhood, no negotiation or agreement is required in its further dissemination. Negotiation on every hop is suitable for interest-based dissemination but not for general interest alerts, for whose verification we rely only on devices in the incident area.\\
Figure \ref{map} details the alert process described in this section. Pedestrian A alerts about an event in its surroundings. The alert is first shared with its neighbors in order to be assessed (green arrows). Then, if they agree with the alert content, the alert is disseminated through the whole area using flooding (red arrows).  
Our contribution is twofold: it ensures better alert quality and decreases network load.

\begin{algorithm}
    \caption{\small{CTD-query event detection}}
    \label{alg:query-detection}
    \begin{algorithmic}[1] % The number tells where the line numbering should start
    
        \STATE event detected({$type$}) 
            \IF{proposal does not exist}
                \STATE alert = create alert proposal (type, time = now, location)
                \STATE send to neighbors(alert)
                \STATE wait (assessment time)
                \IF{most answers are positive}
                \STATE disseminate(alert)
                \ENDIF
            \ENDIF
            \STATE save proposal (alert, time = now)
    \end{algorithmic}
\end{algorithm}

\subsection{Collaborative assessment}\label{subsec:proposal}
This paper does not introduce a different dissemination protocol in mobile networks, but a proposal to include collaborative filtering to decide whether to start the dissemination or not. It includes an alert assessment stage before the ad hoc alert dissemination process in order to assess its reliability (Figure \ref{fig:collaborative-assessment}). We consider two different assessment mechanisms:

\begin{figure}
\centering
\begin{minipage}[b]{.5\linewidth}
\centering
\includegraphics[width=0.8\textwidth]{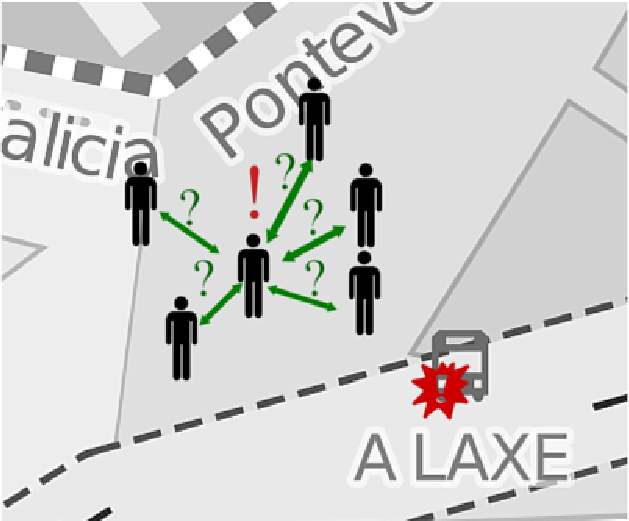}
\caption{Query-based assessment\label{subfig:query_based}}
\end{minipage}%
\begin{minipage}[b]{.5\linewidth}
\centering
\includegraphics[width=0.8\textwidth]{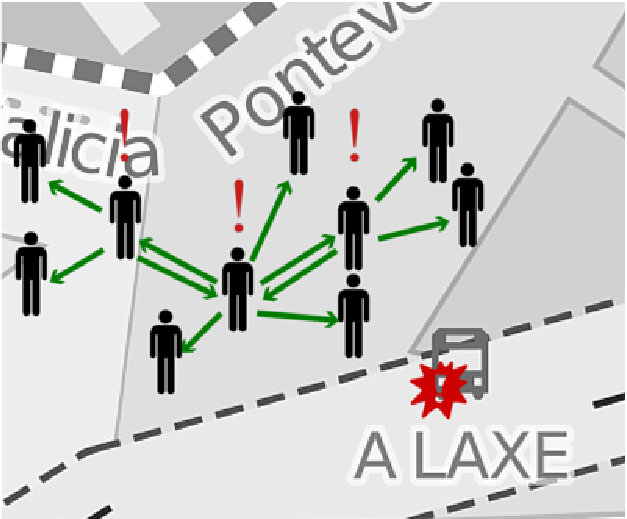}
\caption{Passive assessment\label{subfig:passive}}
\end{minipage}%
\caption{Assessment strategies considered.\label{fig:ctd-strategies}}
\end{figure}

\begin{figure}
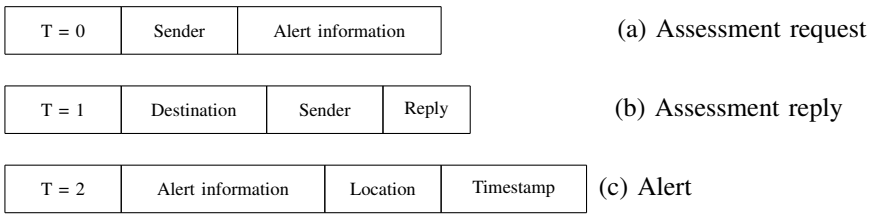

\begin{lrbox}{\bytefieldbox}
%\centering
\begin{bytefield}[bitwidth=1.1em,rightcurly=., rightcurlyspace=62pt]{32}
\begin{rightwordgroup}{ (a) Assessment request\label{subfig:query_request}}
\bitbox{4}{\scriptsize{T = 0}} & \bitbox{4}{\scriptsize{Sender}} & \bitbox{7}{\scriptsize{Alert information}}
\end{rightwordgroup}
\end{bytefield}
\end{lrbox}{\usebox{\bytefieldbox}}
\\

\begin{lrbox}{\bytefieldbox}
%\centering
\begin{bytefield}[bitwidth=1.1em,rightcurly=., rightcurlyspace=50pt]{32}
\begin{rightwordgroup}{ (b) Assessment reply\label{subfig:query_reply}}
\bitbox{4}{\scriptsize{T = 1}} & \bitbox{5}{\scriptsize{Destination}} & \bitbox{4}{\scriptsize{Sender}} & \bitbox{3}{\scriptsize{Reply}}
\end{rightwordgroup}
\end{bytefield}
\end{lrbox}{\usebox{\bytefieldbox}}
\\

\begin{lrbox}{\bytefieldbox}
%\centering
\begin{bytefield}[bitwidth=1.1em,rightcurly=., rightcurlyspace=0pt]{32}
\begin{rightwordgroup}{ (c) Alert\label{subfig:alert}}
\bitbox{4}{\scriptsize{T = 2}} & \bitbox{7}{\scriptsize{Alert information}} & \bitbox{4}{\scriptsize{Location}} & \bitbox{5}{\scriptsize{Timestamp}}
\end{rightwordgroup}
\end{bytefield}
\end{lrbox}{\usebox{\bytefieldbox}}

\caption{Messages employed in CTD-query.\label{fig:query_messages}}
\end{figure}

\begin{itemize}
\item \textit{Query-based collaboratively-triggered dissemination (CTD-query)}. The first node that detects the event sends a request to their neighbors that must decide whether to confirm the alert or not (Figure \ref{fig:ctd-strategies}(\ref{subfig:query_based})). If most of the received replies in a certain time are confirmations, the alert is broadcast. If not, it is discarded. This kind of assessment can be employed in scenarios where nodes' support requires users' intervention or where it is performed autonomously. Users' intervention decreases the probability of errors in event detection but it may be too time-consuming and slow down the system, as users need more time than an automatic system to confirm the alert. Besides, this collaboration depends greatly on users' willingness to cooperate, which is not predictable and makes the system performance variable. Even when replies to queries are autonomously decided by the collaboration system, answering them involves extra work for the nodes as they have to process them and decide if they support them. Query-based operation is shown in Algorithms \ref{alg:query-detection} and \ref{alg:query-received}.
This strategy includes three types of messages: assessment requests, assessment replies and alerts. Their structure is shown in Figure \ref{fig:query_messages}.

\begin{algorithm}
    \caption{\small{CTD-query alert proposal received}}
    \label{alg:query-received}
    \begin{algorithmic}[1] % The number tells where the line numbering should start
        \STATE alert proposal received($alert$) 
                \IF{proposal does not exist}
                \IF{assessment(alert)}
                \STATE reply (positive)
                \ELSE 
                    \STATE reply (negative)
                \ENDIF
                \STATE save proposal (alert, time = now)
                \ENDIF
    \end{algorithmic}
\end{algorithm}

\begin{algorithm}
    \caption{\small{CTD-passive event detection}}
    \label{alg:passive-detection}
    \begin{algorithmic}[1] % The number tells where the line numbering should start
        \STATE event detected($type$)
            \IF{proposal does not exist}
                \STATE alert = create alert proposal (type, time = now, location)
                \STATE send to neighbors(alert)
                 \STATE save proposal (alert, time = now, alert senders = 1)
            \ELSE \IF{proposal already exists AND not disseminated}
                     \STATE increment alert senders (alert)
                       \IF{number of alert senders $>$ (number of neighbors/2)}
                          \STATE disseminate (alert)
                      \ENDIF
                   \ENDIF
            \ENDIF
    \end{algorithmic}
\end{algorithm} 

\item \textit{Passive collaboratively-triggered dissemination (CTD-passive)}. Its operation is illustrated in Figure \ref{fig:ctd-strategies}(\ref{subfig:passive}). Nodes send proposed alerts to their neighbors, which compare them with their own or previously received ones but do not elaborate any reply. Alerts are always proactively decided by every one of the nodes. Thus, the impact of receiving irrelevant messages is minimized. If the node wants to make its own proposal for the same alert, the count is increased and the proposed alert is sent to its neighbors. Nodes keep account of how many of every proposal they have received and, if they have reached the number of required supporters, they disseminate the real alert. The number of necessary supporting nodes depends on the number of neighbors, which is periodically checked. Unlike the query-based method, a node different from the original sender may get enough answers before (because it is getting them from different neighbors) so it will be the one that starts the actual dissemination. It may happen that several nodes reach this amount of messages at the same time but this will not be the same as starting several dissemination processes. Since messages will not include information on the sender, the alert message will be the same (including incident and area)and it will be exactly the same as the ones that may have been sent by other nodes. Nodes remember messages in order not to forward what they have already sent. As a result, dissemination of only one message can be started from several sources. CTD-passive operation is shown in Algorithms \ref{alg:passive-detection} and \ref{alg:passive-received}.\\
In this case, proposed alerts can be launched with or without users' intervention but assessment of received alerts is always automatic. As a result, there exist only two message types: assessment requests and alerts. They have the same structure as the ones shown in Figures \ref{fig:query_messages}(a) and \ref{fig:query_messages}(c).
\end{itemize}

\begin{algorithm}
    \caption{\small{CTD-passive alert proposal received}}
    \label{alg:passive-received}
    \begin{algorithmic}[1] % The number tells where the line numbering should start
        \STATE alert proposal received($type$)
            \IF{proposal does not exist}
                \STATE alert = create alert proposal (type, time = now, location)
                 \STATE save proposal (alert, time = now, alert senders = 1)
            \ELSE \IF{proposal already exists AND it has not been disseminated}
                     \STATE increment alert senders (alert)
                       \IF{number of alert senders $>$ (number of neighbors/2)}
                          \STATE disseminate (alert)
                      \ENDIF
                   \ENDIF
            \ENDIF
    \end{algorithmic}
\end{algorithm}

\subsection{Message processing}
In order to implement collaborative assessment, it is necessary to establish when several alerts can be considered the same. Message processing needs to be made simple, adapted to mobile devices' computing capacity and energy constraints. As a result, complex data mining methods, like the ones used in social networks data processing, are discarded. Furthermore, using simple lightweight messages eases its processing while requiring small bandwidth for transmission.\\ Alert message format will determine how to compare:
\begin{itemize}
\item Free text. Messages can include any text, up to a limited size. This allows the greatest freedom for incidents to be described. However, it only involves good accuracy when the message is both created and assessed directly by users. Topic extraction using text analysis tools could be employed but it complicates message processing and accuracy may not be acceptable.
\item Categorized messages. The system provides a list of incidents that can be subject of alert. Alert description consists only of the incident code, which results in easier message processing. Categories must be narrow enough to be able to give significant information about certain incidents but broad enough not to make the detection process too complex. The list of possible categories can be adapted to different systems' purposes.  
\end{itemize}
Out of the above possibilities, we have chosen categorized messages for our implementation in order to use lightweight message packets that can be automatically processed easily. Selected categories, focused on urban areas, include: traffic, public transport, crowds, crime, nature-related incidents and street shows.

\begin{figure}[h]
\centering
\begin{minipage}[b]{.3\textwidth}
\includegraphics[width=1\textwidth]{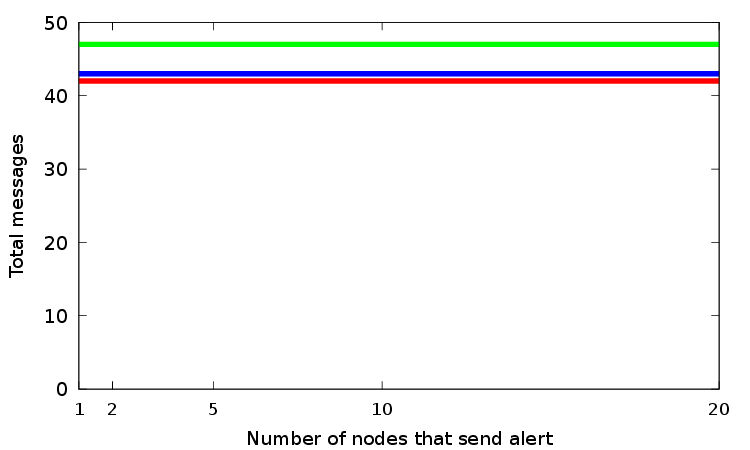}
\caption{100 devices\label{subfig:messages-100}}
\end{minipage}
%\hspace{1mm}
\begin{minipage}[b]{.3\textwidth}
\includegraphics[width=1\textwidth]{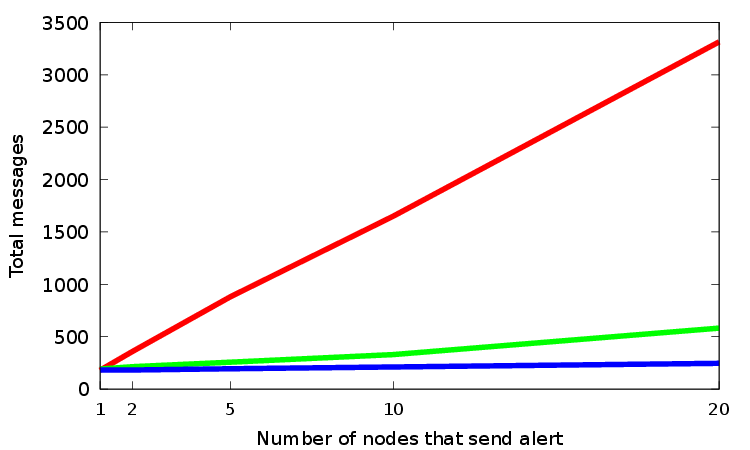}
\caption{200 devices\label{subfig:messages-200}}
\end{minipage}
%\hspace{1mm}
\begin{minipage}[b]{.3\textwidth}
\includegraphics[width=1\textwidth]{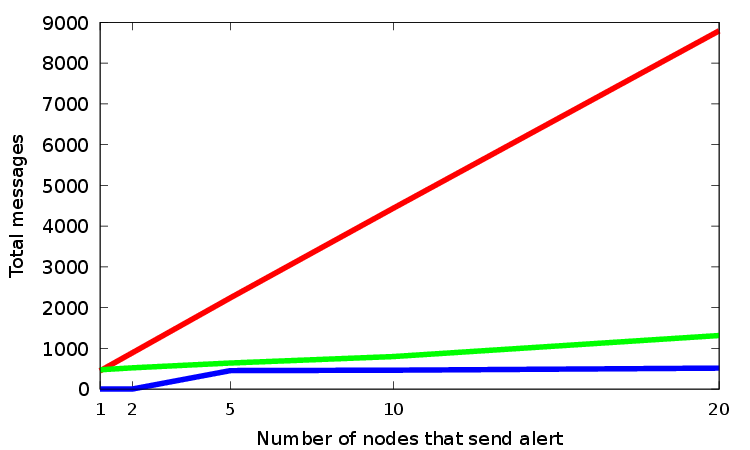}
\caption{500 devices\label{subfig:messages-500}}
\end{minipage}
%\hspace{1mm}
\begin{minipage}[b]{.3\textwidth}
\includegraphics[width=1\textwidth]{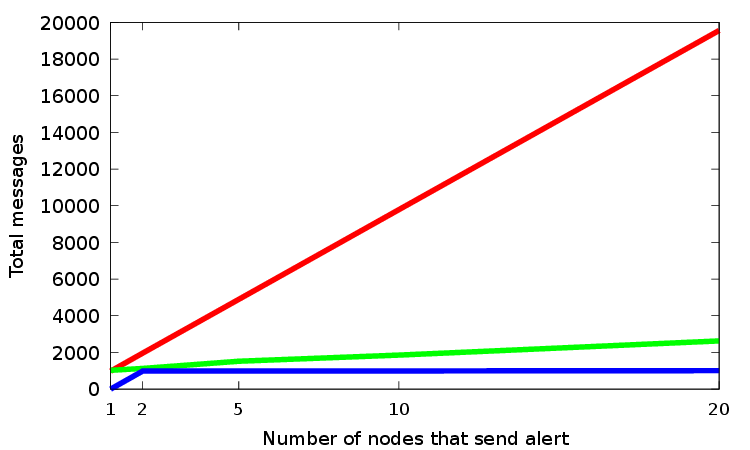}
\caption{1000 devices\label{subfig:messages-1000}}
\end{minipage}
\begin{minipage}[b]{.3\textwidth}
\includegraphics[width=1\textwidth]{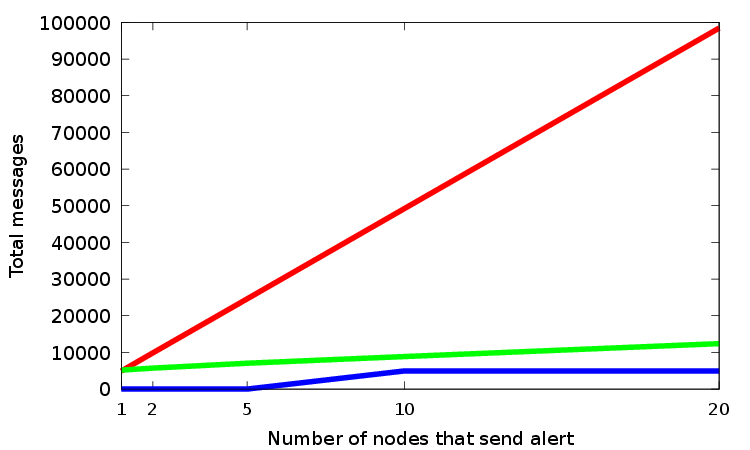}
\caption{5000 devices\label{subfig:messages-5000}}
\end{minipage}
\begin{minipage}[b]{.25\textwidth}
\includegraphics[width=1\textwidth]{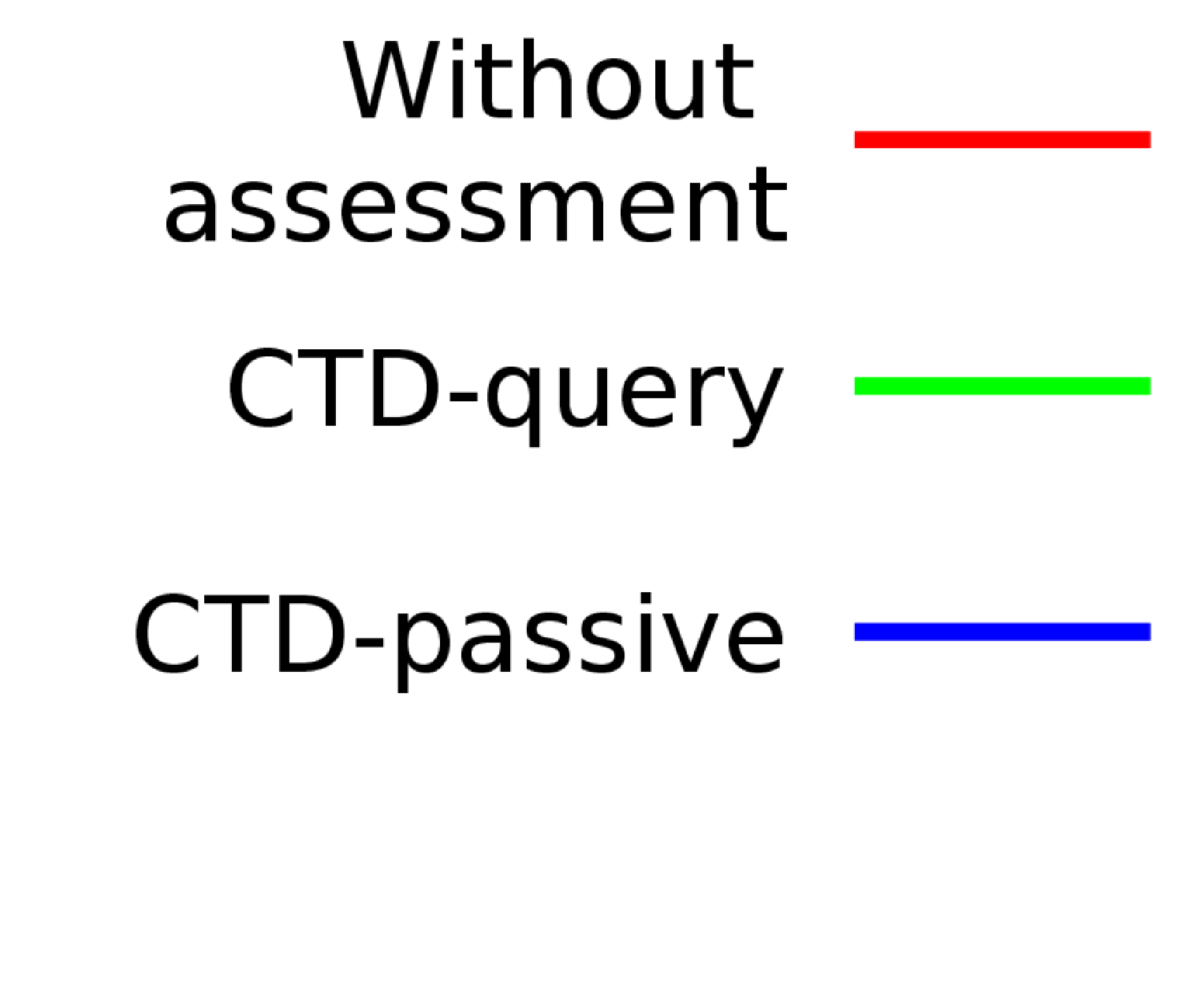}
\end{minipage}

\caption{Total sent messages according to number of devices in the area.\label{fig:messages}}
\end{figure}

\section{Simulation}\label{sec:simulation} 
We have simulated the behavior of our proposal in urban simulations in order to compare our approach with an alert dissemination process without assessment.

\subsection{Network Model}
We consider a set of nodes, randomly placed in an area and moving according to random routes. Their positions and trajectories are always consistent with the areas available for pedestrians (pedestrian streets, pedestrian crossings, sidewalks) in a real map scenario. Nodes are equipped with a short-range communication technology that enables communication with their peers.\\
One or several nodes are chosen to be alert senders, that is to say, to start a dissemination process of the same alert. Alert senders are always physically close, as we model a situation where they all detect the same local event. Every alert sender starts its own dissemination process but not at the same time than the others, there is a time interval (\textit{T}) between their alerts.\\
In CTD-query, neighbors reply to the assessment request as soon as they receive it, without users' intervention. The alert sender waits for replies for a certain time, \textit{W}.
In CTD-passive, there exists a hello mechanism for one-hop neighbor count. This count is employed to calculate how many nodes need to back up an assessment request before being disseminated. The alert dissemination process is triggered if more than half the neighbors support the alert.\\
Assessment requests and alerts are broadcast while assessment replies are sent directly to the alert sender. We assume reliable broadcast. 
For simplicity, we have chosen a controlled flooding mechanism for alert dissemination. Nodes track every alert they receive in a certain time window (\textit{S}) in order not to approve or forward the same twice. Moreover, in CTD-passive, nodes store assessment request together with a count of have many neighbors have backed it up. We assume unlimited storage on the nodes.

\subsection{Urban scenario}
In order to simulate a real-life urban scenario, we have chosen the city center of Vigo (Spain). The chosen region, shown in Figure \ref{map}, comprises an area of 1.46 km$^2$ and includes commercial streets and Vigo's old town. As a result, most of the streets are either pedestrian-only or include wide sidewalks, which makes the area suitable for pedestrian simulation. 
\begin{figure}[h]
\centering
\begin{minipage}[b]{.3\textwidth}
\includegraphics[width=1\textwidth]{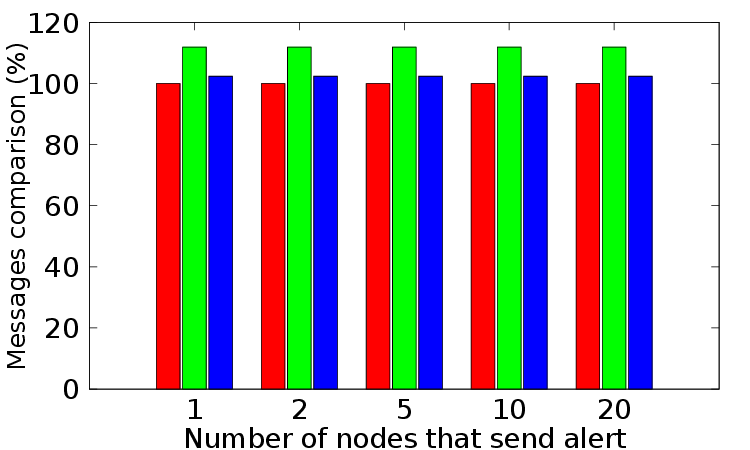}
\caption{100 devices\label{subfig:messages-decrease-100}}
\end{minipage}
%\hspace{1mm}
\begin{minipage}[b]{.3\textwidth}
\includegraphics[width=1\textwidth]{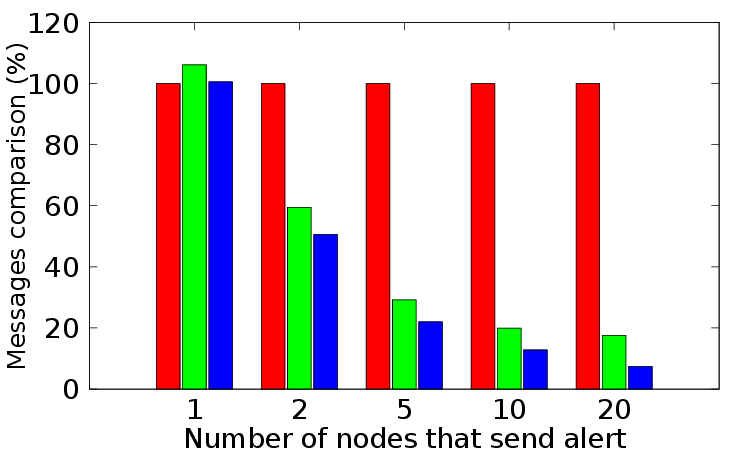}
\caption{200 devices\label{subfig:messages-decrease-200}}
\end{minipage}
%\hspace{1mm}
\begin{minipage}[b]{.3\textwidth}
\includegraphics[width=1\textwidth]{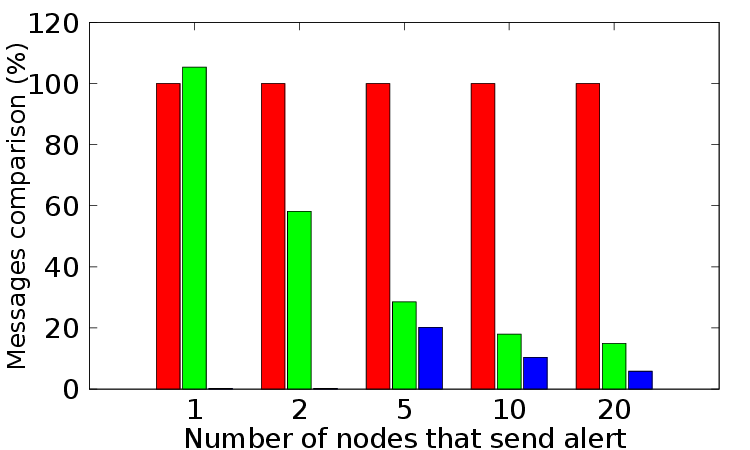}
\caption{500 devices\label{subfig:messages-decrease-500}}
\end{minipage}
%\hspace{1mm}
\begin{minipage}[b]{.3\textwidth}
\includegraphics[width=1\textwidth]{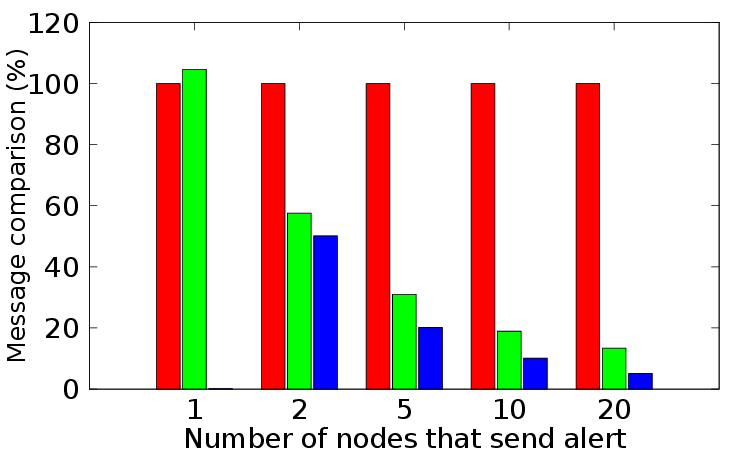}
\caption{1000 devices\label{subfig:messages-decrease-1000}}
\end{minipage}
\begin{minipage}[b]{.3\textwidth}
\includegraphics[width=1\textwidth]{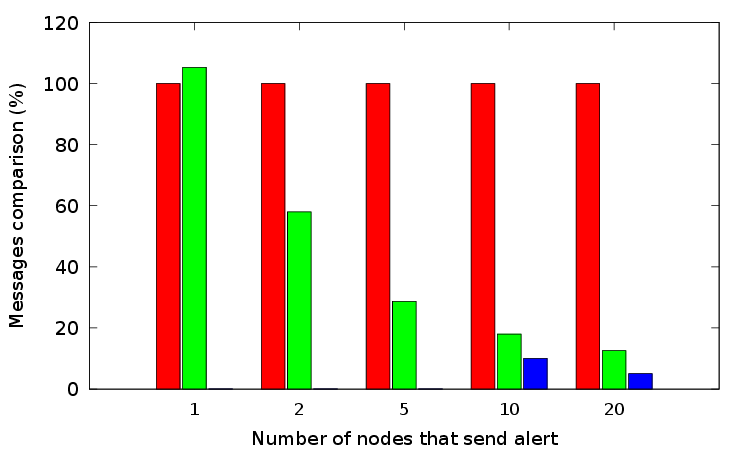}
\caption{5000 devices\label{subfig:messages-decrease-5000}}
\end{minipage}
\begin{minipage}[b]{.25\textwidth}
\includegraphics[width=1\textwidth]{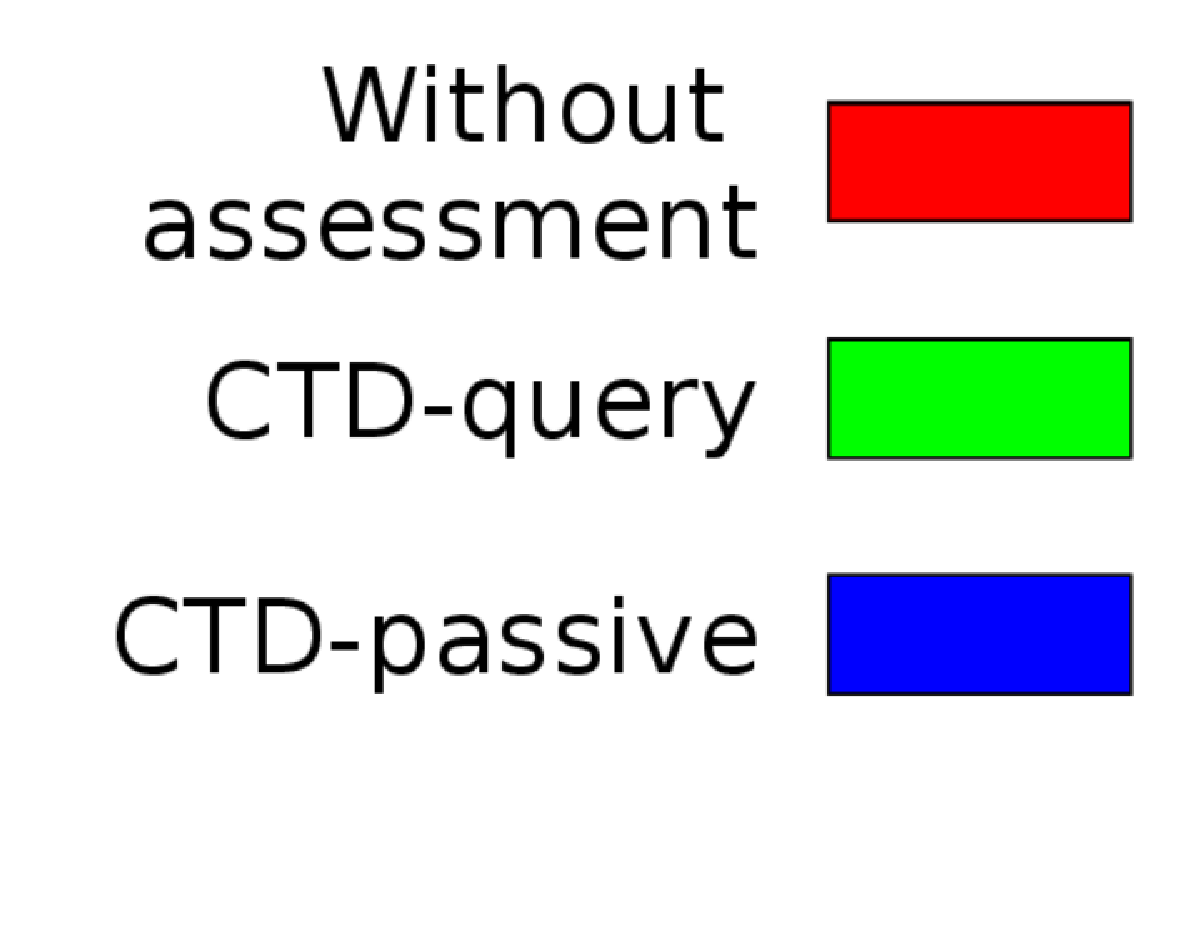}
\end{minipage}

\caption{Percentage of messages sent in relation to the dissemination without assessment.\label{fig:message-decrease}}
\end{figure}

\subsection{Simulation Procedure and Parameters}
We have exported the selected area from OpenStreetMap \cite{OpenStreetMap2015} and populated it with random pedestrian routes with SUMO urban mobility simulator \cite{Krajzewicz2012}. Then, we have exported pedestrian traces to ns-3 network simulator \cite{ns3}, where we have implemented the assessment strategies and carried out the simulation. We have compared them with an alert detection system without assessment in situations where several neighbor devices detect the same incident and alert about it.
\begin{figure}[h]
\centering
\begin{minipage}[b]{.3\textwidth}
\includegraphics[width=1\textwidth]{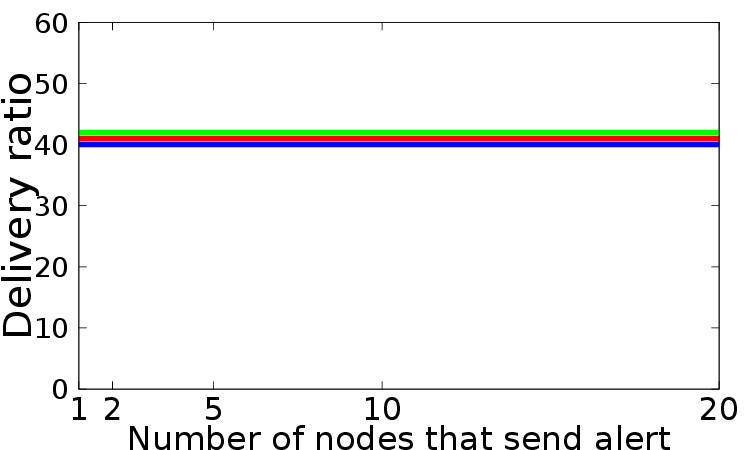}
\caption{100 devices\label{subfig:delivery-ratio-100}}
\end{minipage}
%\hspace{1mm}
\begin{minipage}[b]{.3\textwidth}
\includegraphics[width=1\textwidth]{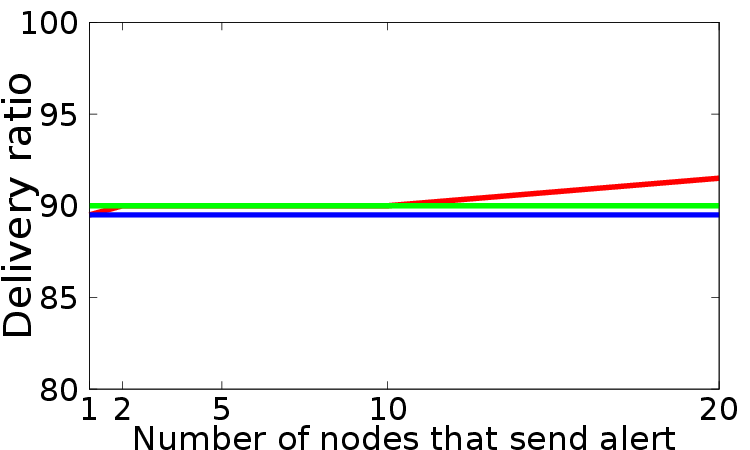}
\caption{200 devices\label{subfig:delivery-ratio-200}}
\end{minipage}
%\hspace{1mm}
\begin{minipage}[b]{.3\textwidth}
\includegraphics[width=1\textwidth]{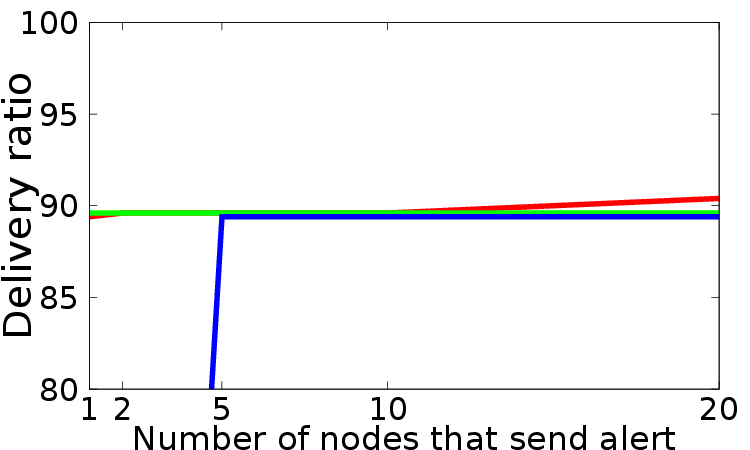}
\caption{500 devices\label{subfig:delivery-ratio-500}}
\end{minipage}
%\hspace{1mm}
\begin{minipage}[b]{.3\textwidth}
\includegraphics[width=1\textwidth]{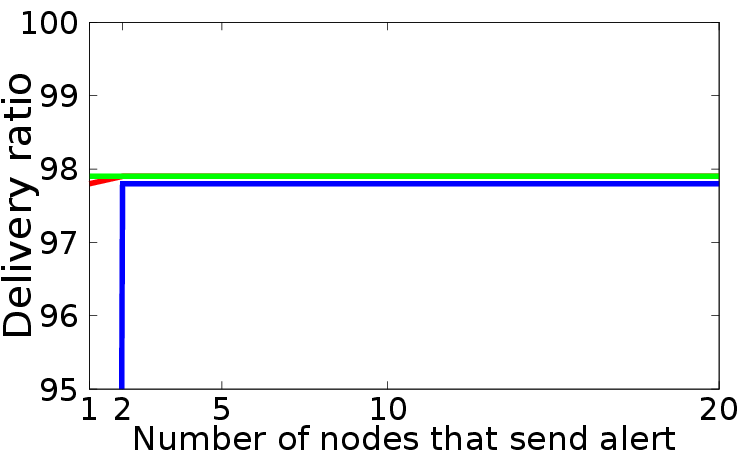}
\caption{1000 devices\label{subfig:delivery-ratio-1000}}
\end{minipage}
\begin{minipage}[b]{.3\textwidth}
\includegraphics[width=1\textwidth]{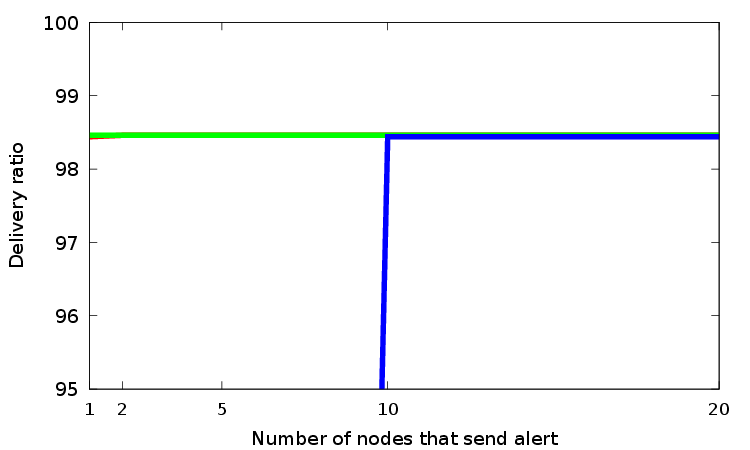}
\caption{5000 devices\label{subfig:delivery-ratio-5000}}
\end{minipage}
\begin{minipage}[b]{.25\textwidth}
\includegraphics[width=1\textwidth]{images/legend-1}
\end{minipage}

\caption{Percentage of nodes in the area that receive the alert.\label{fig:delivery-ratio}}
\end{figure}

We have implemented the two assessment strategies proposed in Section \ref{sec:approach}: CTD-query and CTD-passive. The transmission technology chosen for the implementation was ad-hoc WiFi. Reasons for this election include its availability in current smartphones and its transmission range ($\simeq$ 100m), which makes it appropriate for urban communications.\\
We have carried out simulations with 100, 200, 500, 1000 and 5000 pedestrians in the area in order to assess the collaborative assessment behavior with different device density. In each one of those, we have simulated situations where 1,2, 5, 10 or 20 nodes alert about the incident. These and other parameters related to the simulation can be found in Table \ref{table:simulation-parameters}.

\begin{table}
\centering
\caption{Simulation environment parameters}\label{table:simulation-parameters}
\begin{tabular}{| l | l |}
    
    \hline
    Communication &  Technology: WiFi ad hoc mode \\
    & Connection type: Direct \\
    & Connection pattern: Random \\ \hline
    Environment & Number of nodes: 100,200,500,1000,5000\\
    & Alert senders: 1,2,5,10,20\\
    & Simulation duration: 1h\\
    & Simulation area: 1.46km$^2$\\ \hline
    Mobility & SUMO Random Trips\\ 
     & Speed: 0-1.2 m/s\\ \hline
    CTD parameters & T: 1s\\
     & W: 100ms\\
     & S: 30min\\ \hline
\end{tabular}

\end{table}

\begin{figure}[h]
\centering
\begin{minipage}[b]{.3\textwidth}
\includegraphics[width=1\textwidth]{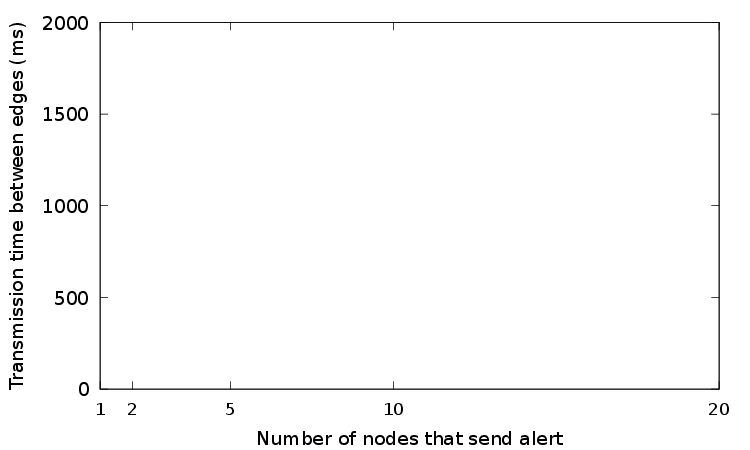}
\caption{100 devices\label{subfig:far-delay-100}}
\end{minipage}
%\hspace{1mm}
\begin{minipage}[b]{.3\textwidth}
\includegraphics[width=1\textwidth]{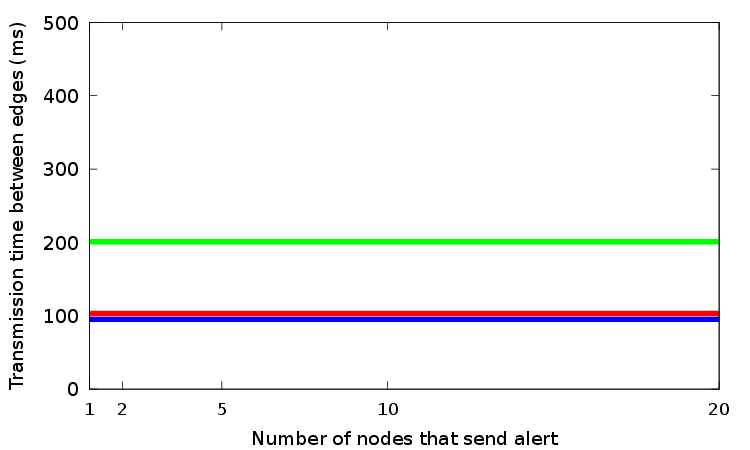}
\caption{200 devices\label{subfig:far-delay-200}}
\end{minipage}
%\hspace{1mm}
\begin{minipage}[b]{.3\textwidth}
\includegraphics[width=1\textwidth]{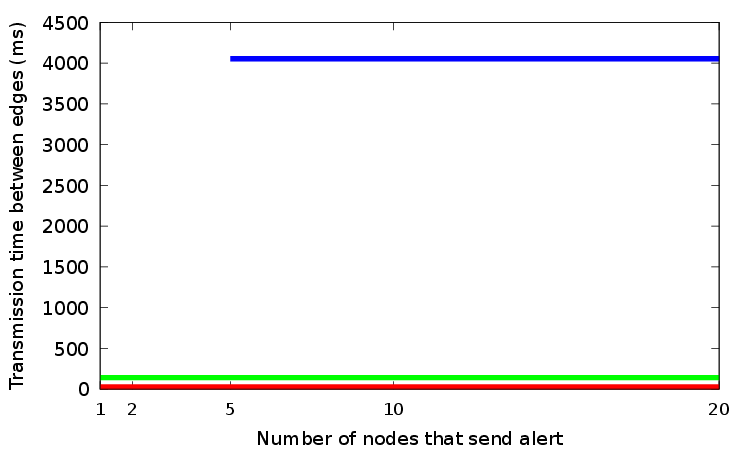}
\caption{500 devices\label{subfig:far-delay-500}}
\end{minipage}
%\hspace{1mm}
\begin{minipage}[b]{.3\textwidth}
\includegraphics[width=1\textwidth]{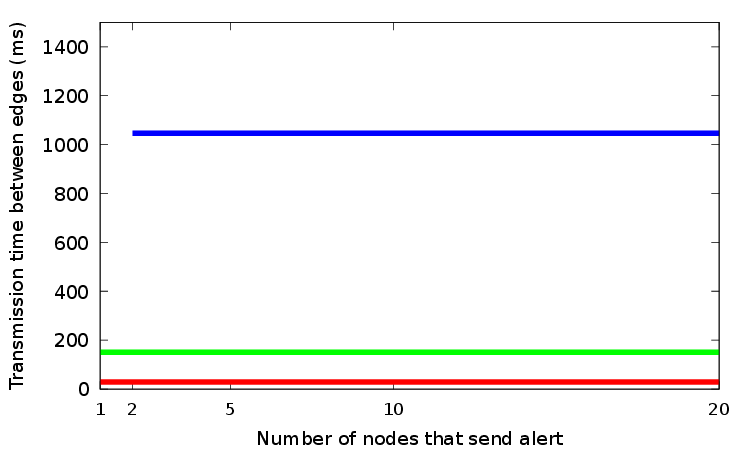}
\caption{1000 devices\label{subfig:far-delay-1000}}
\end{minipage}
\begin{minipage}[b]{.3\textwidth}
\includegraphics[width=1\textwidth]{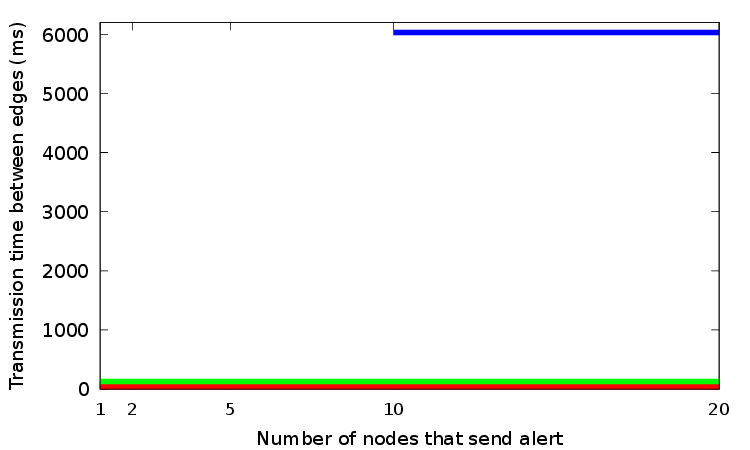}
\caption{5000 devices\label{subfig:far-delay-5000}}
\end{minipage}
\begin{minipage}[b]{.25\textwidth}
\includegraphics[width=1\textwidth]{images/legend-1}
\end{minipage}

\caption{Transmission time between opposite edges of the area.\label{fig:far-delay}}
\end{figure}

The following network parameters were considered as performance metrics:
\begin{itemize}
\item Message efficiency. We have measured the total number of messages that are necessary for an alert dissemination, including assessment requests, replies and disseminated alerts. In addition to the direct comparison of the absolute number of messages, we provide graphs that depict the percentage of messages employed in relation to the amount of messages required in a simulation where no assessment strategy is employed. Simulations without assessment require 100\% of messages and results with our proposed strategies are shown as a function of these. This representation offers a more visual approach and makes it easier to detect  the same percentage of transmission reduction in simulations with different pedestrian density and, therefore, with large differences in the number of messages.
\item Delivery ratio. We have measured the percentage of nodes in the area that receive the alert. Due to the randomness of the positioning and the characteristics of the urban area, some pedestrians are placed on the edges and cannot be reached. For this reason, delivery ratio never achieves 100\%.
\item Transmission delay between opposite area edges. It is measured as transmission time from the original sender to a device on the opposite edge of the considered area. We have used pedestrians marked as A and B in Figure \ref{map}.
\item Transmission time to cover the alert area. It is measured as the transmission time from the original sender to the last node that receives the alert. Although this is a parameter that should be minimized, it is reasonable to have a larger delay if delivery ratio is increased.
\end{itemize}
\begin{figure}[h]
\centering
\begin{minipage}[b]{.3\textwidth}
\includegraphics[width=1\textwidth]{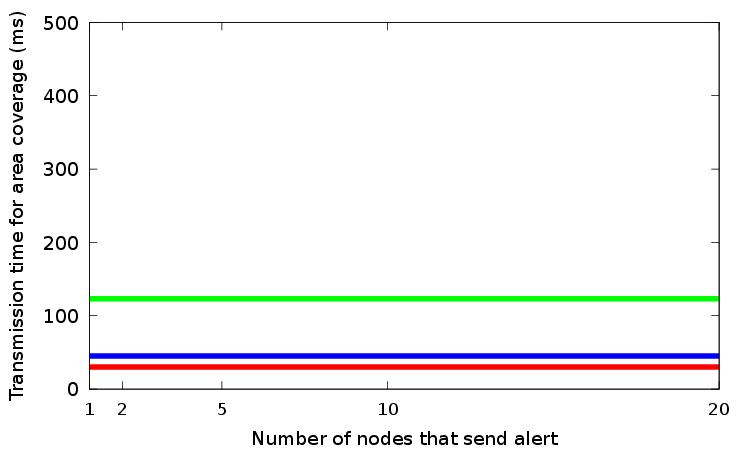}
\caption{100 devices\label{subfig:overall-delay-100}}
\end{minipage}
%\hspace{1mm}
\begin{minipage}[b]{.3\textwidth}
\includegraphics[width=1\textwidth]{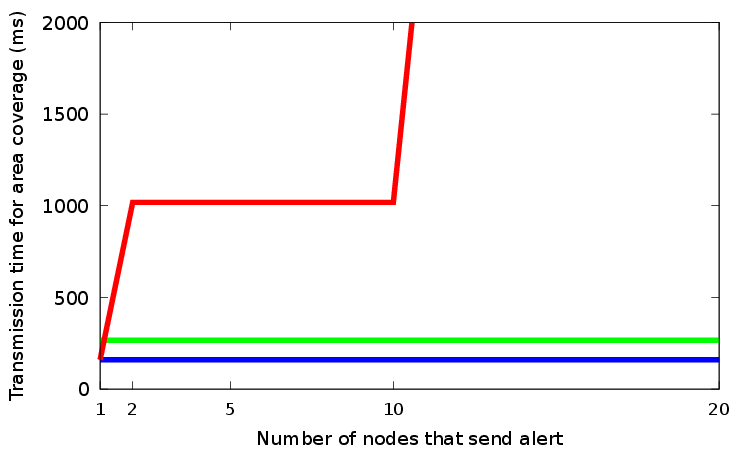}
\caption{200 devices\label{subfig:overall-delay-200}}
\end{minipage}
%\hspace{1mm}
\begin{minipage}[b]{.3\textwidth}
\includegraphics[width=1\textwidth]{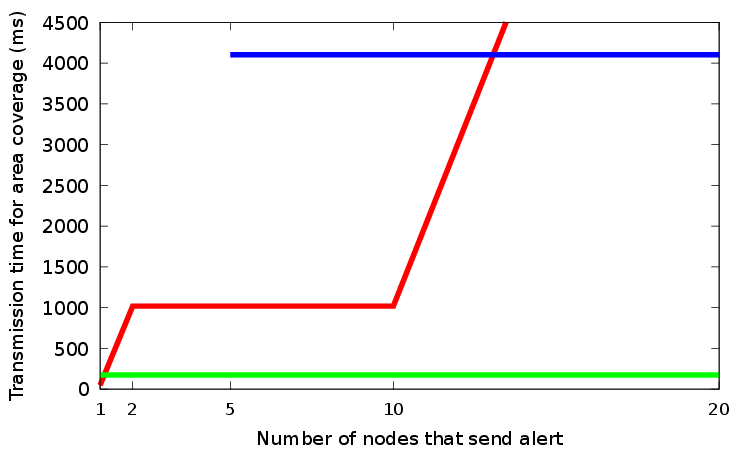}
\caption{500 devices\label{subfig:overall-delay-500}}
\end{minipage}
%\hspace{1mm}
\begin{minipage}[b]{.3\textwidth}
\includegraphics[width=1\textwidth]{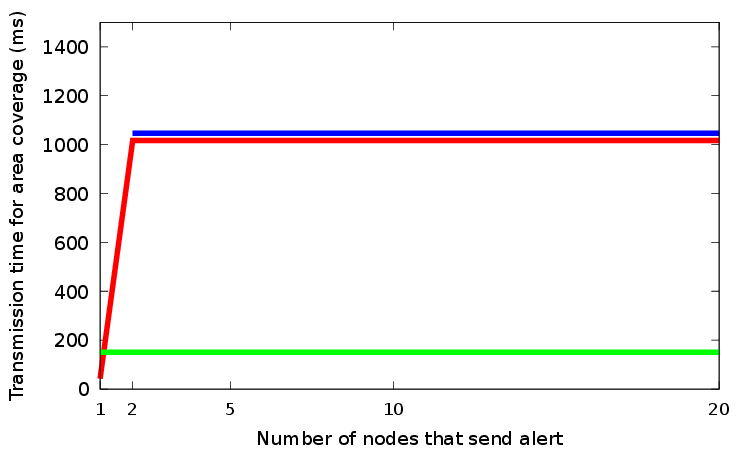}
\caption{1000 devices\label{subfig:overall-delay-1000}}
\end{minipage}
\begin{minipage}[b]{.3\textwidth}
\includegraphics[width=1\textwidth]{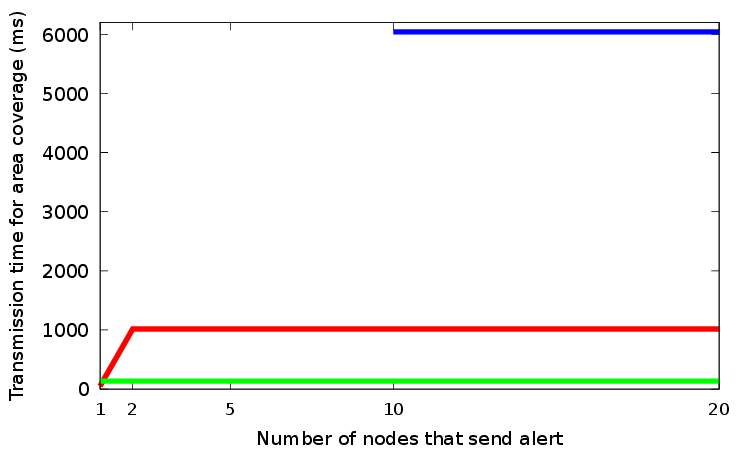}
\caption{5000 devices\label{subfig:overall-delay-5000}}
\end{minipage}
\begin{minipage}[b]{.25\textwidth}
\includegraphics[width=1\textwidth]{images/legend-1}
\end{minipage}

\caption{Transmission time to cover the area.\label{fig:overall-delay}}
\end{figure}

\subsection{Modeling nodes' collaboration}
In the first simulations, we have assumed all the alerts are relevant and therefore, that they will be all backed up by neighbors. However, it is also possible that there are selfish nodes that refuse to collaborate or that some nodes do not approve the alert content. In order to analyze those scenarios, we have performed simulations where nodes accept the alert proposal with a certain probability ($p_a$). We have focused on scenarios where the alert dissemination is likely to be triggered, where alerts are indeed relevant ($p_a = 0.9$ and $p_a = 0.75$) but part of the nodes disagrees or refuses to collaborate. Lower $p_a$ values correspond to non-collaborative scenarios or scenarios with untrue or irrelevant alerts, where the dissemination would not be triggered. As a result, they cannot be object of a network analysis that enables comparison with the first simulations.
We have performed these simulations for 1, 2, 5, 10 and 20 alert senders out of 100, 200, 500 and 1000 devices.

\section{Evaluation}\label{sec:evaluation}
We have concluded that both assessment mechanisms outperform dissemination without assessment as long as more than one device alerts about the incident and there is enough device density. 100 devices have been proved to be too few to cover the considered area. In this simulation opposite end is never reached (Figure \ref{fig:far-delay}(\ref{subfig:far-delay-100})) and delivery ratio is below 60\% (Figure \ref{fig:delivery-ratio}(\ref{subfig:delivery-ratio-100})). Distance between nodes prevents further dissemination while maintaining the quantity of messages sent constant. Moreover, assessment stage messages make CTD mechanisms more inefficient than dissemination without assessment, as shown in Figure \ref{fig:messages}(\ref{subfig:messages-100}) and \ref{fig:message-decrease}(\ref{subfig:messages-decrease-100}). With low densities (100-200 devices) nodes have only few close neighbors, CTD-passive dissemination is then easily triggered, even with only one node that sends the alert. The number of neighbors agreeing nodes required increases with device density. In the 5000 devices case, higher device density makes it necessary that at least 10 devices send the alert in order to trigger the CTD-passive dissemination.\\
From 200 devices on, we can observe a clear trend on the system behavior. For one alert sender message efficiency without assessment outperforms CTD-query (Figures \ref{fig:messages} and\ref{fig:message-decrease}) as it does not require any extra messages apart from the dissemination ones. For two alert messages, CTD-query and CTD-passive (when triggered) require about half of the messages employed without assessment. As the number of the nodes that send the alert increases, so does the difference between the dissemination without assessment and the collaborative assessment approaches. With 20 alert senders, both require less than 20\% of the messages the former employs. And that happens with little difference in the delivery ratio, which is about 90\% with 200 and 500 devices and around 98\% for 1000 and 5000 (Figure \ref{fig:delivery-ratio}). \\
Regarding dissemination times, transmission to end node is ensured by enough device density and delay is constant for every mechanism, as shown in Figure \ref{fig:far-delay}.
We have included Figure \ref{fig:far-delay}(\ref{subfig:far-delay-100}) for consistency even though no transmission to the opposite edge were possible with such a low device density. CTD-query alert is always 100ms later than dissemination without assessment since this is the time alert senders wait for its peers' assessment. For CTD-passive, time to start dissemination is not fixed but dependent on the number of neighbors that are required to back up the alert. As the alert interval has been set to one second in these simulations, when two alerts are required time to start dissemination will be two seconds, three seconds when three alerts are required and so on. As a result, it increases as node density does. Time to reach the opposite edge node in the 200 devices case is similar to the one employed when no assessment is involved (Figure \ref{fig:far-delay}(\ref{subfig:far-delay-200})) as the alert is early disseminated, as explained in the first paragraph. Anyway, higher dissemination time resulted from simulation was 6 seconds (Figure \ref{fig:overall-delay}(\ref{subfig:overall-delay-500})), which is still suitable for an alert system. 

Figure \ref{fig:overall-delay} shows how the node selected in the opposite edge is not the most difficult to be reached. This node (pedestrian B in Figure \ref{map}) is indeed placed on the further area of the map but the path to it goes through usually-populated wide streets. Due to the random positioning of the pedestrians, some of them have been placed in smaller routes or passages with little device density around them. As a consequence, the possibility to reach them results from opportunistic node movements (of them or their peers) and may be missed if the dissemination start time is different. That is the reason that explains a higher delivery ratio in some cases of dissemination without assessment, where area coverage time is much larger (Figures \ref{fig:delivery-ratio}(\ref{subfig:delivery-ratio-200}) and \ref{fig:overall-delay}(\ref{subfig:overall-delay-200})). Taking into account the extra messages and time consumed, the cost of reaching those nodes is high. 

\begin{figure}
\centering
\begin{minipage}[b]{.4\textwidth}
\includegraphics[width=1\textwidth]{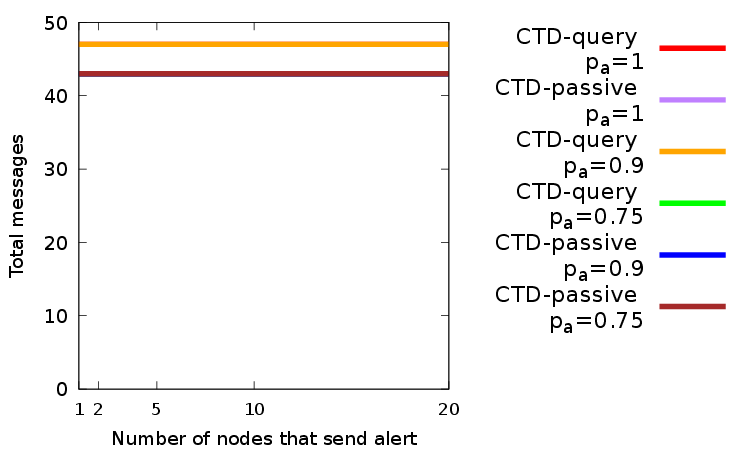}
\caption{100 devices\label{subfig:messages-100-prob}}
\end{minipage}
%\hspace{1mm}
\begin{minipage}[b]{.4\textwidth}
\includegraphics[width=1\textwidth]{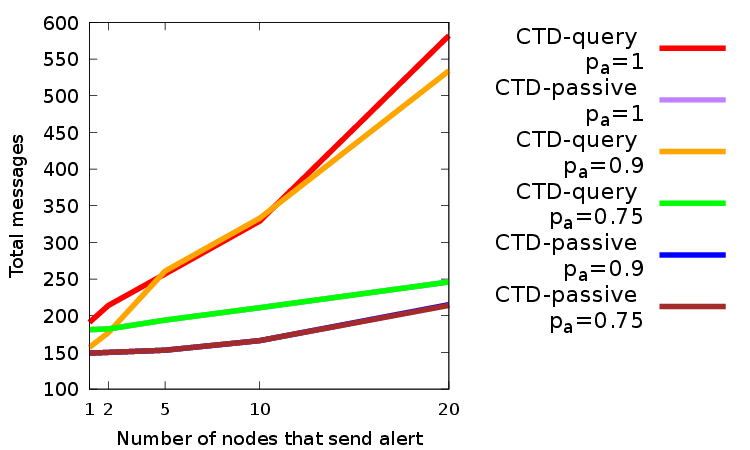}
\caption{200 devices\label{subfig:messages-200-prob}}
\end{minipage}
%\hspace{1mm}
\begin{minipage}[b]{.4\textwidth}
\includegraphics[width=1\textwidth]{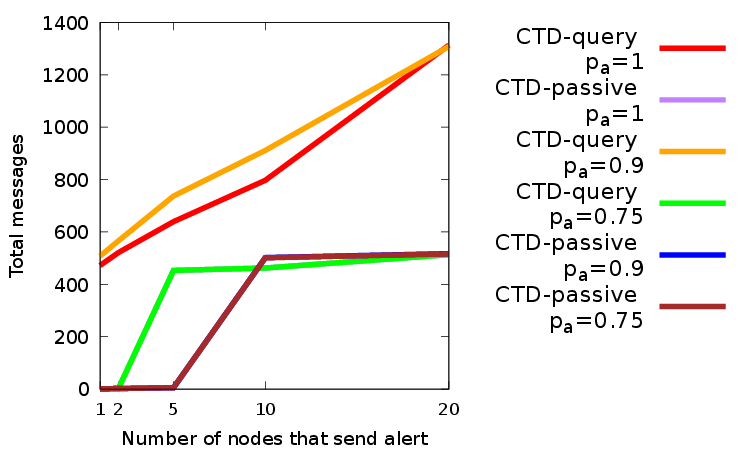}
\caption{500 devices\label{subfig:messages-500-prob}}
\end{minipage}
%\hspace{1mm}
\begin{minipage}[b]{.4\textwidth}
\includegraphics[width=1\textwidth]{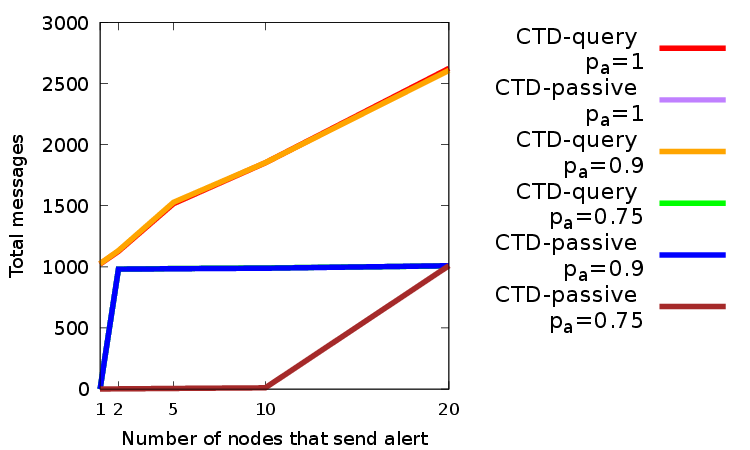}
\caption{1000 devices\label{subfig:messages-1000-prob}}
\end{minipage}
\caption{Comparison of the number of messages as a function of the assessment probability.\label{fig:messages-prob}}
\end{figure}

\begin{figure}
\centering
\begin{minipage}[b]{.4\textwidth}
\includegraphics[width=1\textwidth]{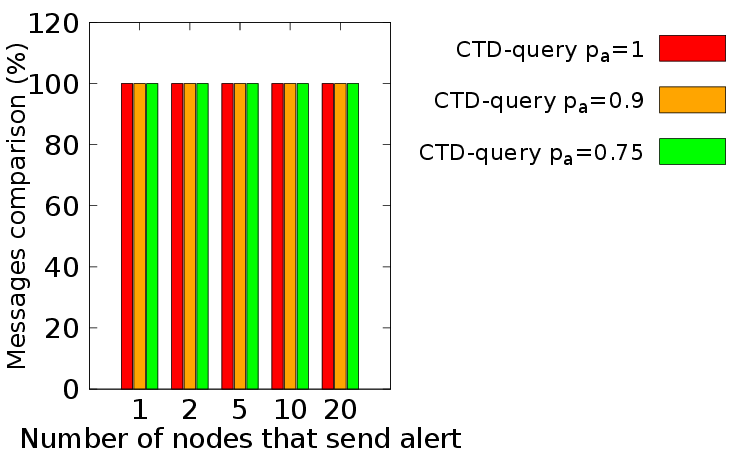}
\caption{100 devices\label{subfig:messages-decrease-100-prob-query}}
\end{minipage}
%\hspace{1mm}
\begin{minipage}[b]{.4\textwidth}
\includegraphics[width=1\textwidth]{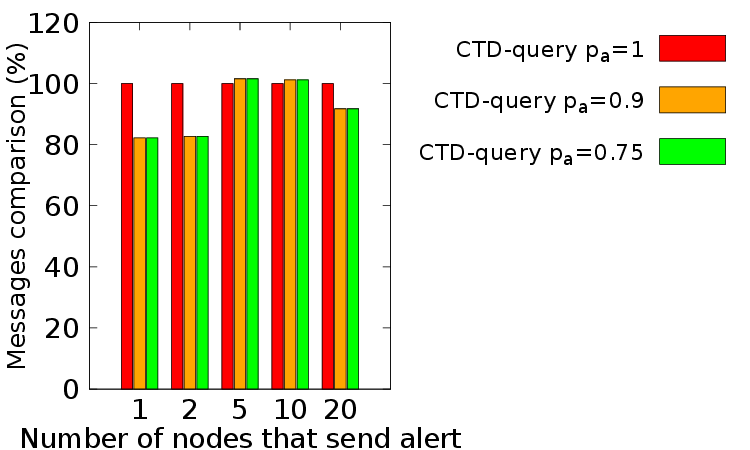}
\caption{200 devices\label{subfig:messages-decrease-200-prob-query}}
\end{minipage}
%\hspace{1mm}
\begin{minipage}[b]{.4\textwidth}
\includegraphics[width=1\textwidth]{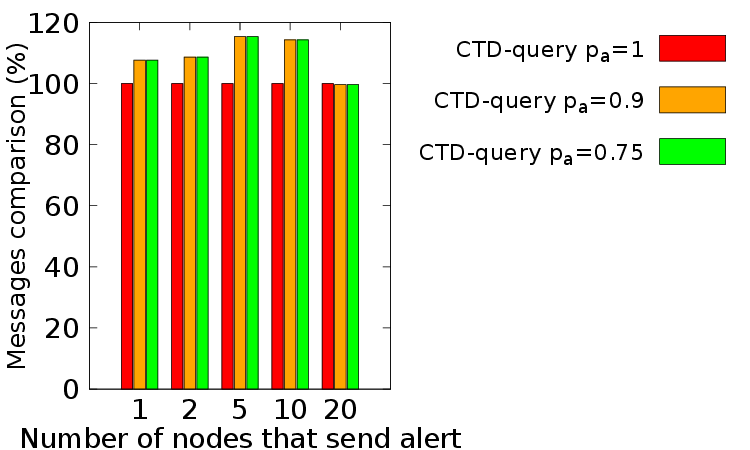}
\caption{500 devices\label{subfig:messages-decrease-500-prob-query}}
\end{minipage}
%\hspace{1mm}
\begin{minipage}[b]{.4\textwidth}
\includegraphics[width=1\textwidth]{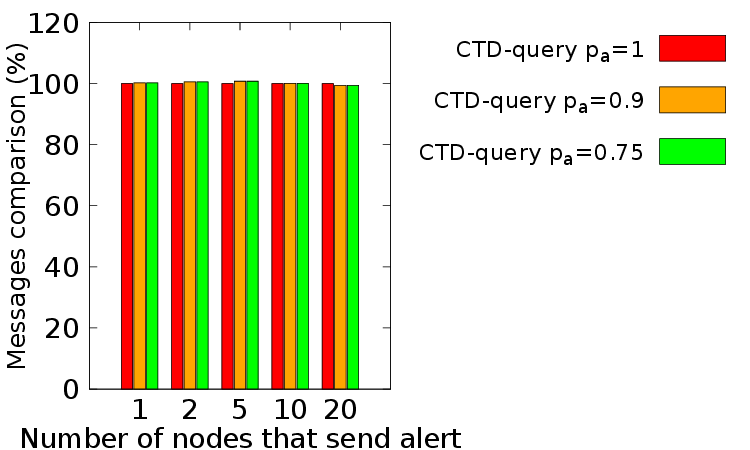}
\caption{1000 devices\label{subfig:messages-decrease-1000-prob-query}}
\end{minipage}

\caption{Comparison of messages sent as a function of the assessment probability (CTD-query).\label{fig:message-decrease-query-prob}}
\end{figure}

\begin{figure}
\centering
\begin{minipage}[b]{.4\textwidth}
\includegraphics[width=1\textwidth]{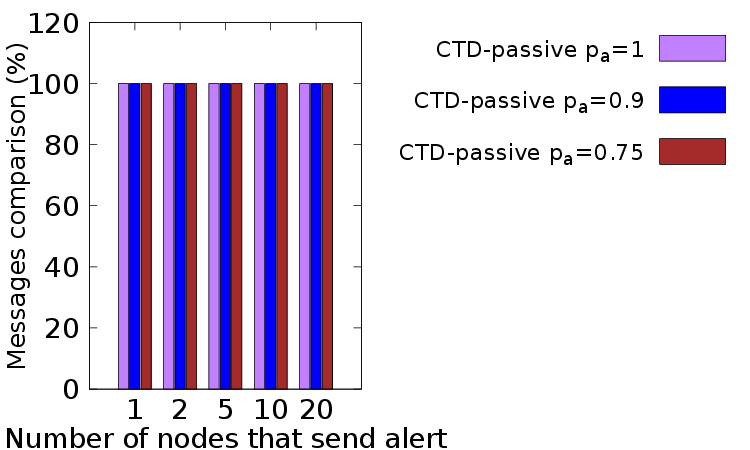}
\caption{100 devices\label{subfig:messages-decrease-100-prob-passive}}
\end{minipage}
%\hspace{1mm}
\begin{minipage}[b]{.4\textwidth}
\includegraphics[width=1\textwidth]{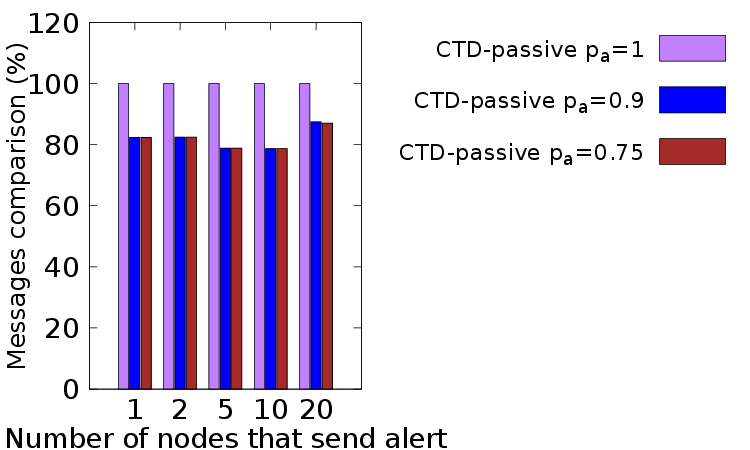}
\caption{200 devices\label{subfig:messages-decrease-200-prob-passive}}
\end{minipage}
%\hspace{1mm}
\begin{minipage}[b]{.4\textwidth}
\includegraphics[width=1\textwidth]{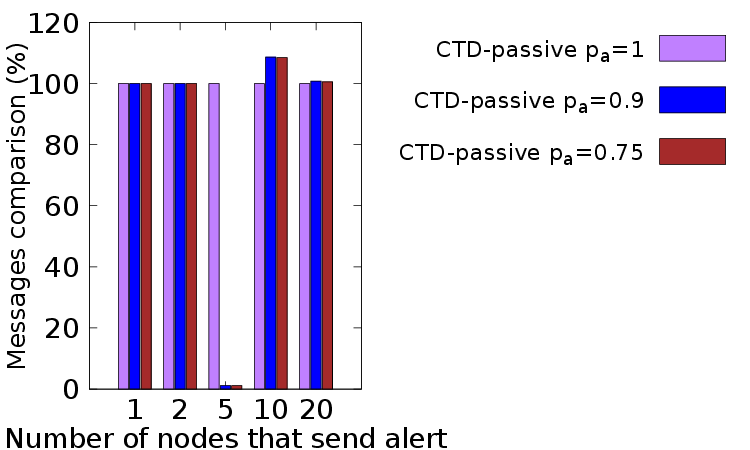}
\caption{500 devices\label{subfig:messages-decrease-500-prob-passive}}
\end{minipage}
%\hspace{1mm}
\begin{minipage}[b]{.4\textwidth}
\includegraphics[width=1\textwidth]{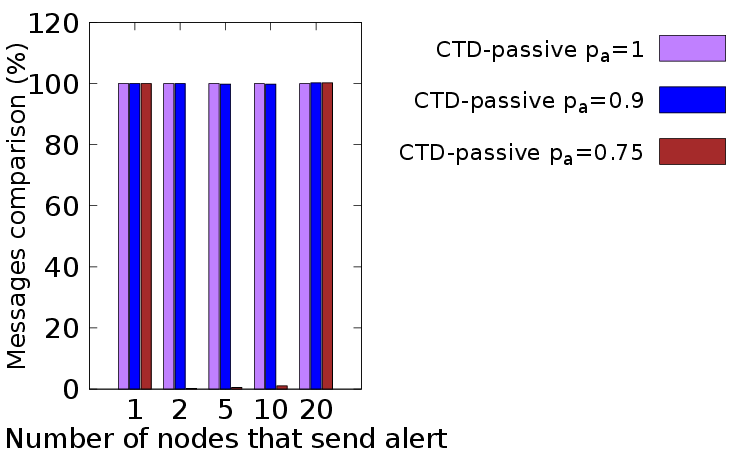}
\caption{1000 devices\label{subfig:messages-decrease-1000-prob-passive}}
\end{minipage}
\caption{Comparison of messages sent as a function of the assessment probability (CTD-passive).\label{fig:message-decrease-passive-prob}}
\end{figure}

\begin{figure}
\centering
\begin{minipage}[b]{.4\textwidth}
\includegraphics[width=1\textwidth]{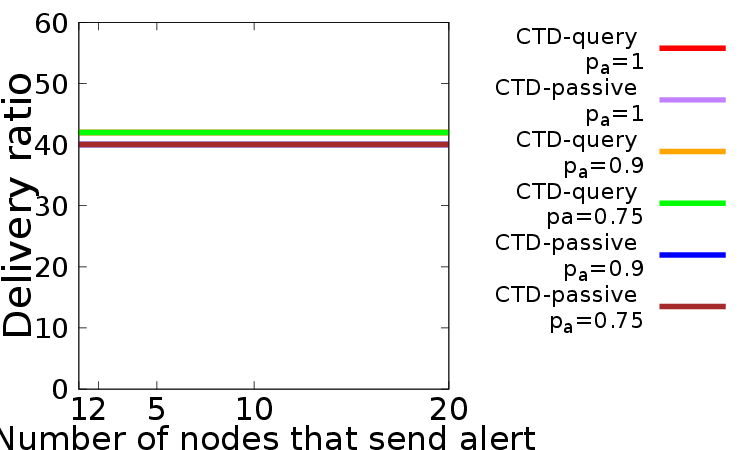}
\caption{100 devices\label{subfig:delivery-ratio-100-prob}}
\end{minipage}
%\hspace{1mm}
\begin{minipage}[b]{.4\textwidth}
\includegraphics[width=1\textwidth]{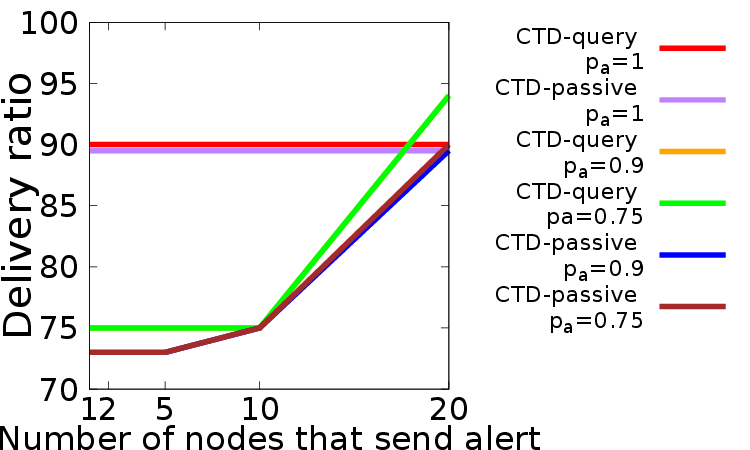}
\caption{200 devices\label{subfig:delivery-ratio-200-prob}}
\end{minipage}
%\hspace{1mm}
\begin{minipage}[b]{.4\textwidth}
\includegraphics[width=1\textwidth]{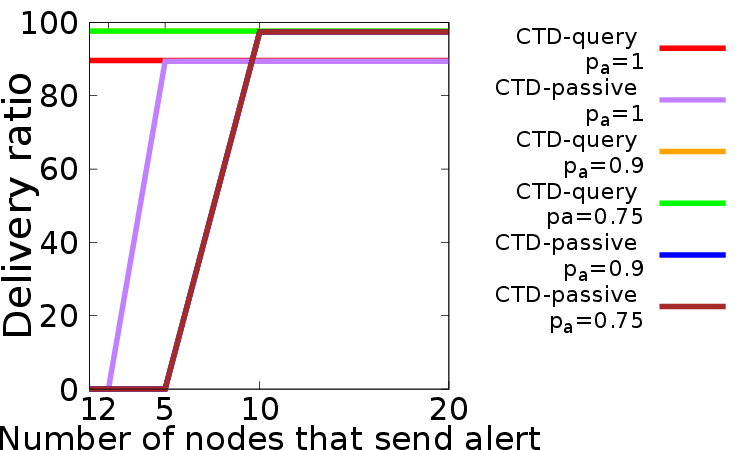}
\caption{500 devices\label{subfig:delivery-ratio-500-prob}}
\end{minipage}
%\hspace{1mm}
\begin{minipage}[b]{.4\textwidth}
\includegraphics[width=1\textwidth]{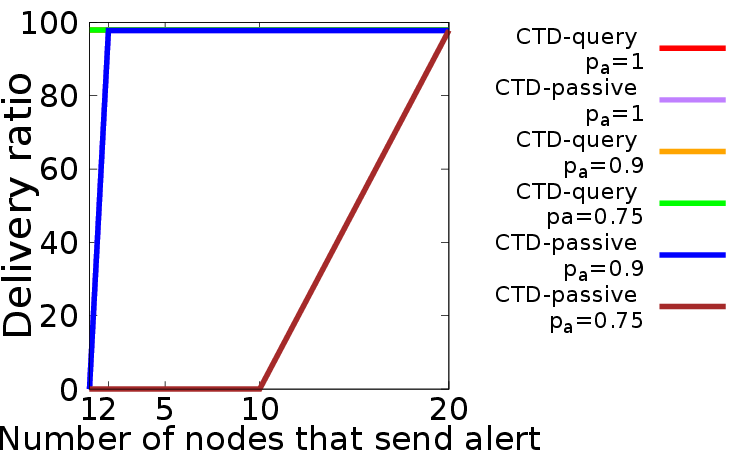}
\caption{1000 devices\label{subfig:delivery-ratio-1000-prob}}
\end{minipage}
\caption{Percentage of nodes in the area that receive the alert as a function of the assessment probability.\label{fig:delivery-ratio-prob}}
\end{figure}

\begin{figure}
\centering
\begin{minipage}[b]{.4\textwidth}
\includegraphics[width=1\textwidth]{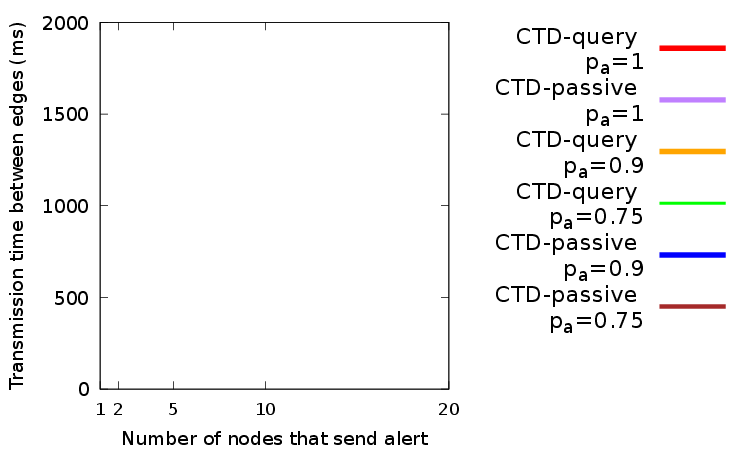}
\caption{100 devices\label{subfig:far-delay-100-prob}}
\end{minipage}
%\hspace{1mm}
\begin{minipage}[b]{.4\textwidth}
\includegraphics[width=1\textwidth]{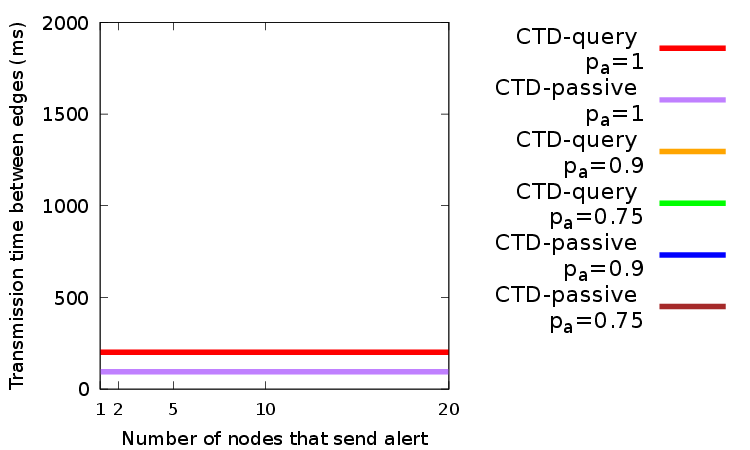}
\caption{200 devices\label{subfig:far-delay-200-prob}}
\end{minipage}
%\hspace{1mm}
\begin{minipage}[b]{.4\textwidth}
\includegraphics[width=1\textwidth]{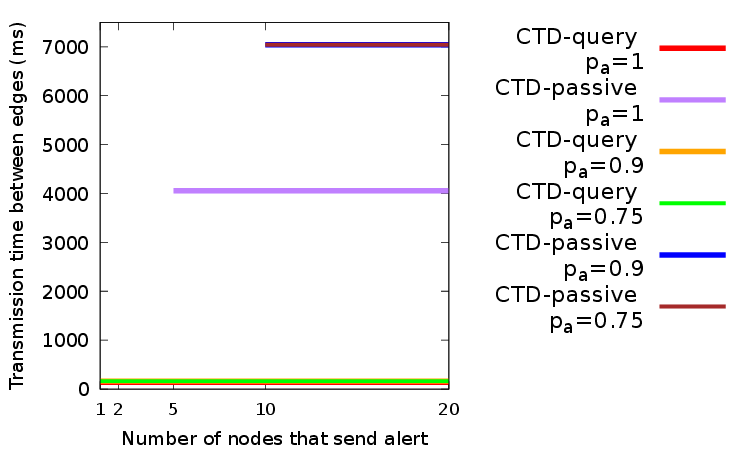}
\caption{500 devices\label{subfig:far-delay-500-prob}}
\end{minipage}
%\hspace{1mm}
\begin{minipage}[b]{.4\textwidth}
\includegraphics[width=1\textwidth]{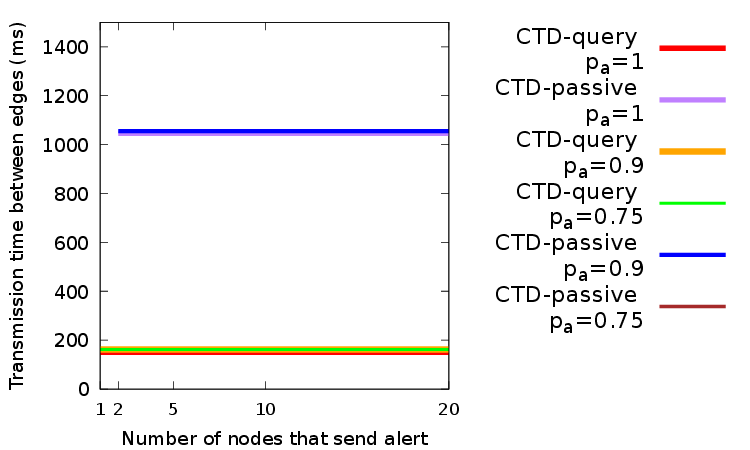}
\caption{1000 devices\label{subfig:far-delay-1000-prob}}
\end{minipage}
\caption{Transmission time between opposite edges of the area as a function of the assessment probability.\label{fig:far-delay-prob}}
\end{figure}

\begin{figure}
\centering
\begin{minipage}[b]{.4\textwidth}
\includegraphics[width=1\textwidth]{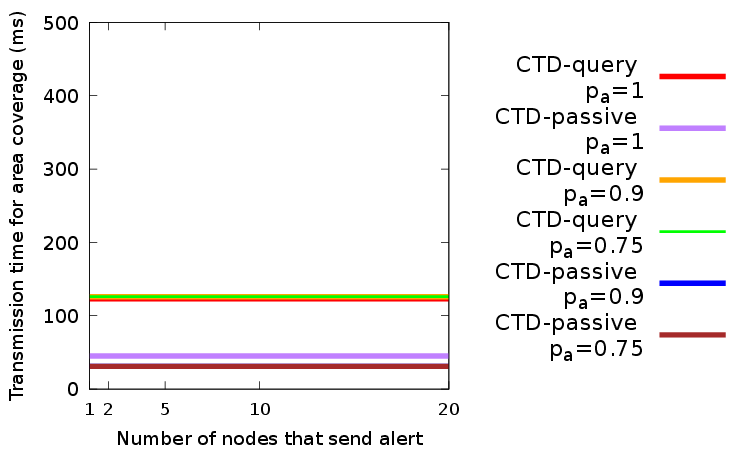}
\caption{100 devices\label{subfig:overall-delay-100-prob}}
\end{minipage}
%\hspace{1mm}
\begin{minipage}[b]{.4\textwidth}
\includegraphics[width=1\textwidth]{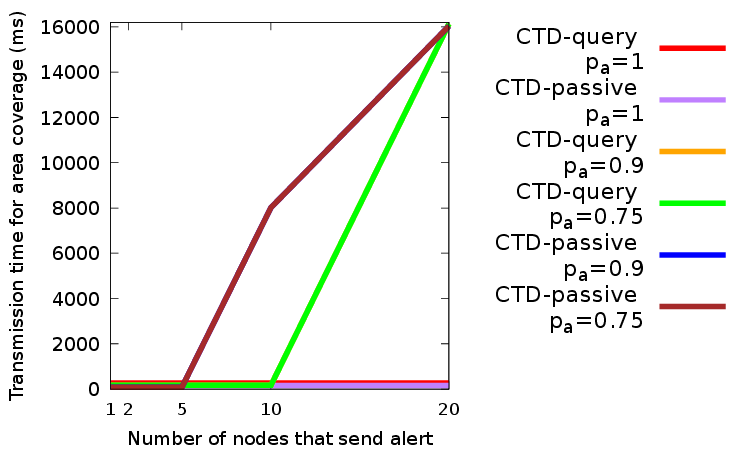}
\caption{200 devices\label{subfig:overall-delay-200-prob}}
\end{minipage}
%\hspace{1mm}
\begin{minipage}[b]{.4\textwidth}
\includegraphics[width=1\textwidth]{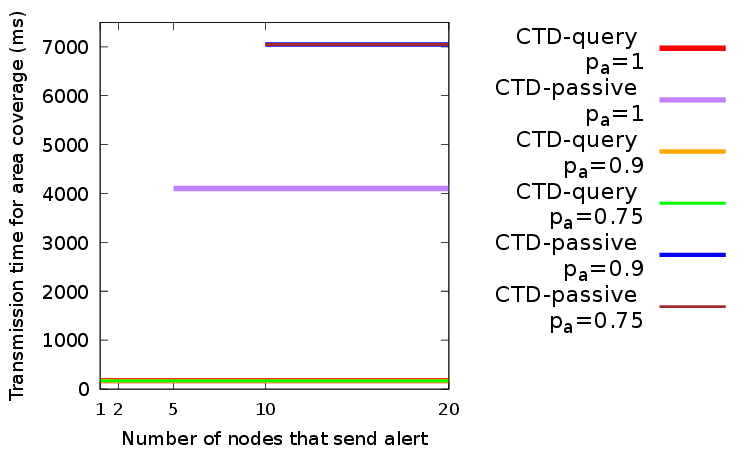}
\caption{500 devices\label{subfig:overall-delay-500-prob}}
\end{minipage}
%\hspace{1mm}
\begin{minipage}[b]{.4\textwidth}
\includegraphics[width=1\textwidth]{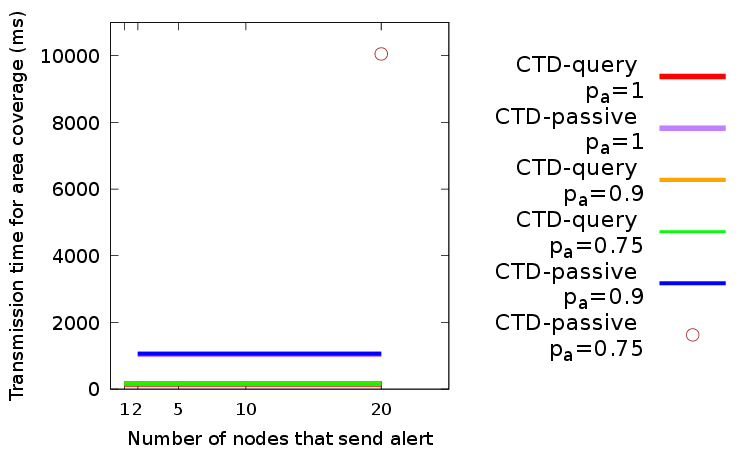}
\caption{1000 devices\label{subfig:overall-delay-1000-prob}}
\end{minipage}
\caption{Transmission time to cover the area as a function of the assessment probability.\label{fig:overall-delay-prob}}
\end{figure}

\subsection{Modeling nodes' collaboration}
We have analyzed the same network parameters in scenarios where nodes may refuse to accept the alert with a low probability. We have performed the simulations of our two assessment mechanisms: CTD-query and CTD-passive with high assessment probabilities ($p_a = 0.9$ and $p_a = 0.75$)and we have compared them with the results detailed in the previous section, where $p_a = 1$. \\
The results show that the alert system is still working although a higher device density is required to ensure a good delivery ratio. In the 100-nodes scenario, the results are only slightly different. Due to the low density, nodes have very few neighbors and, being the $p_a$ values so high, their behavior is unlikely to vary. As a result, there are no changes in the messages sent (Figures \ref{fig:messages-prob}(\ref{subfig:messages-100-prob}), \ref{fig:message-decrease-query-prob}(\ref{subfig:messages-decrease-100-prob-query}) and \ref{fig:message-decrease-passive-prob}(\ref{subfig:messages-decrease-100-prob-passive})), nor in the delivery ratio (Figure \ref{fig:delivery-ratio-prob}(\ref{subfig:delivery-ratio-100-prob})), nor in the far delay (Figure \ref{fig:far-delay-prob}(\ref{subfig:far-delay-100-prob})). Besides, changes on the overall delay are minimal (Figure \ref{fig:overall-delay-prob}(\ref{subfig:overall-delay-100-prob})). \\
In the most populated scenario, changes in CTD-query results are also small but for the opposite reason. Since nodes have a large number of neighbors, they are likely to receive enough replies even when some of them refuse to collaborate. Therefore, the number of sent messages is maintained (Figures \ref{fig:messages-prob}(\ref{subfig:messages-1000-prob}) and \ref{fig:message-decrease-query-prob}(\ref{subfig:messages-decrease-1000-prob-query})), as well as the delivery ratio (Figure \ref{fig:delivery-ratio-prob}(\ref{subfig:delivery-ratio-1000-prob})) and the transmission times. However, results differ for CTD-passive, when $p_a = 0.75$. In this situation, dissemination is not triggered with less than 20 alert senders. For 20 alert senders, the alert system matches the performance of previous simulations in every parameter, except for the transmission time to cover the whole area  (Figure \ref{fig:overall-delay-prob}(\ref{subfig:overall-delay-1000-prob})), which significantly increases because of the effect of the mobility on the different transmission times.\\ 
While in the lowest and highest densities the results are almost maintained, they differ in the medium ones. In the 200-nodes scenario, the delivery ratio decreases with the number of nodes that send the alert (Figure \ref{fig:delivery-ratio-prob}(\ref{subfig:delivery-ratio-200-prob})). At the same time, the amount of sent messages is reduced. However, as the alert dissemination is always triggered, it is never too low. To reach delivery ratio values similar to the ones obtained with $p_a = 1$, it is necessary that 20 nodes send the alert, both for CTD-query and CTD-passive. The alert never reaches the opposite edge node, not even in this case. Therefore, there are not any new data in Figure \ref{fig:far-delay-prob}(\ref{subfig:far-delay-200-prob}), which is the same as Figure \ref{fig:far-delay}(\ref{subfig:far-delay-200}) and has been included for consistency, like the graph depicting the 100-nodes case. As well as in the 1000-nodes case, the overall transmission delay significantly increases in some simulations (Figure \ref{fig:overall-delay-prob}(\ref{subfig:overall-delay-200-prob})). Since this increase does not happen together with an increment in the delivery ratio or in the quantity of messages, it indicates that it took more time to reach the same number of devices. This can only be caused by devices' mobility in the area and the different timing of the dissemination trigger.\\
Mobility also influences the dissemination in the 500-nodes case. In this scenario, the different timing of the dissemination provoked by the assessment probability enables more useful encounters between nodes in the area. This leads to an increased delivery ratio (Figure \ref{fig:delivery-ratio-prob}(\ref{subfig:delivery-ratio-500-prob})), which requires a larger amount of messages when CTD-query in employed (Figure \ref{fig:message-decrease-query-prob}(\ref{subfig:messages-decrease-500-prob-query})) and when CTD-passive is triggered (Figure \ref{fig:message-decrease-passive-prob}(\ref{subfig:messages-decrease-500-prob-passive})). This takes place from two alert senders on when $p_a = 0.9$ and from five senders on when $p_a = 0.75$. For fewer alert senders, the amount of messages matches the one in scenarios where $p_a = 1$ because dissemination was not triggered in those either. The transmission time to cover the area grows (Figure \ref{fig:overall-delay-prob}(\ref{subfig:overall-delay-500-prob})), although not as much as in the 200 and 1000-nodes cases. \\
In conclusion, the lack of collaboration from a limited group of nodes does not seriously harm the alert system performance when there are at least 500 devices in the area. For lower densities, the results decline unless the number of alert senders increases.

\subsection{Discussion}
To the best of our knowledge, we are the first to design a collaborative assessment mechanism for ad hoc alert systems that includes the assessment of every created alert. Moreover, we are also the first to analyze how to reduce network impact considering alert content.\\
Systems based on content-aware prioritization \cite{Uddin2011} deal with duplicates but not irrelevant information being shared. They aim at disseminating as many alerts as possible and, only when there are space constraints the similarity between them is considered. As long as there is space, duplicate alerts can still be disseminated. As a result, the whole available space could be filled with instances of the same alert. In contrast, we focus on the assessment itself and on the control of the network load. Everything that has been found worthy to be disseminated is sent, what has not is ignored. Moreover, content-aware prioritization includes a processing overhead in every node, which we limit to the ones in the assessment area. Since the prioritization goal is to maximize the number of alerts, their evaluation is targeted at this parameter. Our objective is different and, therefore, so is our evaluation.\\
Opportunist collaboration systems are often evaluated in terms of energy expenditure \cite{Lee2012, Shi2011}. Although we have not specifically studied this issue in our system, the reduction of network load implies less processing overhead on the nodes and should consequently translate into a lower energy expenditure. \\
Nodes' collaboration is not only critical in content-based assessment but also in every ad hoc communication strategy. Cacciapuoti et al. \cite{Cacciapuoti2013} have studied the impact of the number of enabling and active nodes, analog to our alert senders and collaborating nodes, in an ad hoc system targeted at emergency communication. However, their approach is different. Nodes are randomly assigned to collaborate or not and maintain the same status during the whole simulation. Our nodes decide to collaborate or not every time they receive an alert to be assessed.  

\section{Conclusions and Future Work} \label{sec:conclusion}
We have presented two collaborative assessment proposals for ad hoc alert systems in smart cities. We have detailed their operation while comparing them with other alert detection and dissemination mechanisms. Finally, we have simulated their behavior and discussed their network performance and the impact of different degrees of node collaboration.\\
To the best of our knowledge, no other work has been presented targeted at collaboratively filtering what is worthy to be disseminated in an ad-hoc alert system. Our proposal has been proved able to reduce network load while maintaining similar delivery ratio. Moreover, it does not require setup cost or negotiations on every step. The quantity of messages employed in the assessment phase has been proved to be small and therefore, the strategy cost is minimal.\\
In this paper, we have studied an urban scenario where incidents are detected by users carrying mobile devices or by devices being carried by users. Nevertheless, this is not the only scene that can benefit from collaborative alert assessment. It can be applied to any situations where mobile sensors communicate ad hoc. The list includes sensor networks, vehicular networks, unmanned aerial vehicles (UAV) networks and will increase exponentially with the rise of connected objects as part of the Internet of Things development. Therefore, our strategies can be employed in many different scenarios that incorporate an alert system such as smart homes, industrial monitoring and intelligent transportation. In addition to alert systems, applications that can employ ad hoc collaborative assessment schemes include intrusion detection and information gathering in architectures where several sensors are covering the same area and need to agree on an only sensed value to contribute to the system.\\
However, our approach can only be applied to alerts that can be also be observed by neighbors. This is enough for most city alerts that can be triggered from a smartphone or other handheld device (traffic jams, queues) but may not be enough in other scenarios, such as vehicular networks, where alerts may inform about breakdowns or other problems in the very vehicle that can only be assessed by itself.\\
Even though our proposal has been proved useful for message verification and network load reduction, it is true that it does not address other concerns related to ad hoc networks, such as selfish nodes. On the contrary, our proposal presents a bigger dependency on neighbors' cooperation, as it is required to trigger the dissemination process. However, this should not be a problem in urban scenarios where device density is high. Based on the number of users' contributions to relevant event detection that can be found on social networks, we believe they will be inclined to cooperate in this kind of system. Our current work focuses on the study of security issues, such as malicious alerts, that can be handled using our collaborative approach.

\bibliography{bibtex.bib}

\begin{thebibliography}{999}
\expandafter\ifx\csname url\endcsname\relax
  \def\url#1{\texttt{#1}}\fi
\expandafter\ifx\csname urlprefix\endcsname\relax\def\urlprefix{URL }\fi
\expandafter\ifx\csname href\endcsname\relax
  \def\href#1#2{#2} \def\path#1{#1}\fi

\bibitem{Campbell2006}
A.~T. Campbell, S.~B. Eisenman, N.~D. Lane, E.~Miluzzo, R.~A. Peterson, {People-centric urban sensing}, WICON '06 (2006) 18--31\href {https://doi.org/10.1145/1234161.1234179} {\path{doi:10.1145/1234161.1234179}}.

\bibitem{Burke2006}
J.~Burke, D.~Estrin, M.~Hansen, A.~Parker, N.~Ramanathan, S.~Reddy, M.~B. Srivastava, {Participatory Sensing}, Center for Embedded Network Sensing. (2006) 1--5.

\bibitem{Guo2014}
B.~Guo, Z.~Yu, X.~Zhou, {From Participatory Sensing to Mobile Crowd Sensing}, in: Pervasive Computing and Communications Workshops (PERCOM Workshops), 2014, pp. 593--598.

\bibitem{Guo2015}
B.~I.~N. Guo, Z.~H.~U. Wang, Z.~Yu, Y.~U. Wang, N.~Y. Yen, R.~Huang, X.~Zhou, {Mobile Crowd Sensing and Computing: The Review of an Emerging Human-Powered Sensing Paradigm}, ACM Computing Surveys 48~(1) (2015) 7.
\newblock \href {https://doi.org/10.1145/2794400} {\path{doi:10.1145/2794400}}.

\bibitem{Zhang2014}
D.~Zhang, Z.~Yu, B.~Guo, Z.~Wang, \href{http://dx.doi.org/10.1007/978-1-4614-8579-7{\_}6}{{Exploiting Personal and Community Context in Mobile Social Networks}}, in: A.~Chin, D.~Zhang (Eds.), Mobile Social Networking: An Innovative Approach, Springer New York, New York, NY, 2014, pp. 109--138.
\newblock \href {https://doi.org/10.1007/978-1-4614-8579-7_6} {\path{doi:10.1007/978-1-4614-8579-7_6}}.
\newline\urlprefix\url{http://dx.doi.org/10.1007/978-1-4614-8579-7{\_}6}

\bibitem{Uddin2011}
M.~Y.~S. Uddin, H.~Wang, F.~Saremi, G.-j. Qi, T.~Abdelzaher, T.~Huang, {PhotoNet: A Similarity-aware Picture Delivery Service for Situation Awareness}, in: Real-Time Systems Symposium (RTSS), 2011 IEEE 32nd, 2011, pp. 1--10.

\bibitem{Cacciapuoti2013}
A.~S. Cacciapuoti, F.~Calabrese, M.~Caleffi, G.~{Di Lorenzo}, L.~Paura, \href{http://dx.doi.org/10.1016/j.pmcj.2012.01.002}{{Human-mobility enabled wireless networks for emergency communications during special events}}, Pervasive and Mobile Computing 9~(4) (2013) 472--483.
\newblock \href {https://doi.org/10.1016/j.pmcj.2012.01.002} {\path{doi:10.1016/j.pmcj.2012.01.002}}.
\newline\urlprefix\url{http://dx.doi.org/10.1016/j.pmcj.2012.01.002}

\bibitem{Sun2003}
B.~Sun, K.~Wu, U.~W. Pooch, {Alert Aggregation in Mobile Ad Hoc Networks}, WiSE'03 (2003) 69--78\href {https://doi.org/10.1145/941311.941323} {\path{doi:10.1145/941311.941323}}.

\bibitem{Castro-Jul2016}
F.~Castro-Jul, R.~P. Diaz-Redondo, A.~Fernandez-Vilas, {Have You Also Seen That? Collaborative Alert Assessment in Ad Hoc Participatory Sensing}, in: Ubiquitous Computing and Ambient Intelligence: 10th International Conference, UCAmI 2016, San Bartolom{\'{e}} de Tirajana, Gran Canaria, Spain, Springer International Publishing., 2016, pp. 125--130.
\newblock \href {https://doi.org/10.1007/978-3-319-48799-1_15} {\path{doi:10.1007/978-3-319-48799-1_15}}.

\bibitem{Fujiki2008}
Y.~Fujiki, K.~Kazakos, C.~Puri, P.~Buddharaju, I.~Pavlidis, \href{http://portal.acm.org.cyber.usask.ca/citation.cfm?id=1371216.1371224{\%}7B{\&}{\%}7Dcoll=ACM{\%}7B{\&}{\%}7Ddl=ACM{\%}7B{\&}{\%}7DCFID=101604612{\%}7B{\&}{\%}7DCFTOKEN=69013139}{{NEAT-o-Games: Blending Physical Activity and Fun in the Daily Routine}}, Computers in Entertainment (CIE) 6~(2) (2008) 1--22.
\newblock \href {https://doi.org/10.1145/1371216.1371224} {\path{doi:10.1145/1371216.1371224}}.
\newline\urlprefix\url{http://portal.acm.org.cyber.usask.ca/citation.cfm?id=1371216.1371224{\%}7B{\&}{\%}7Dcoll=ACM{\%}7B{\&}{\%}7Ddl=ACM{\%}7B{\&}{\%}7DCFID=101604612{\%}7B{\&}{\%}7DCFTOKEN=69013139}

\bibitem{DHondt2013}
E.~D'Hondt, M.~Stevens, A.~Jacobs, \href{http://dx.doi.org/10.1016/j.pmcj.2012.09.002}{{Participatory noise mapping works! An evaluation of participatory sensing as an alternative to standard techniques for environmental monitoring}}, Pervasive and Mobile Computing 9~(5) (2013) 681--694.
\newblock \href {https://doi.org/10.1016/j.pmcj.2012.09.002} {\path{doi:10.1016/j.pmcj.2012.09.002}}.
\newline\urlprefix\url{http://dx.doi.org/10.1016/j.pmcj.2012.09.002}

\bibitem{Sakaki2010}
T.~Sakaki, M.~Okazaki, Y.~Matsuo, {Earthquake shakes Twitter users}, in: WWW '10, ACM Press, 2010, p. 851.
\newblock \href {https://doi.org/10.1145/1772690.1772777} {\path{doi:10.1145/1772690.1772777}}.

\bibitem{Eriksson2008}
J.~Eriksson, L.~Girod, B.~Hull, R.~Newton, S.~Madden, H.~Balakrishnan, {The Pothole Patrol: Using a Mobile Sensor Network for Road Surface Monitoring}, in: Proceedings of the 6th international conference on Mobile systems, applications, and services, 2008, pp. 29--39.

\bibitem{Zhou2012}
P.~Zhou, Y.~Zheng, M.~Li, {How Long to Wait?: Predicting Bus Arrival Time with Mobile Phone based Participatory Sensing}, in: Proceedings of the 10th international conference on Mobile systems, applications, and services, 2012, pp. 379--392.

\bibitem{Imran2015}
M.~Imran, C.~Castillo, F.~Diaz, S.~Vieweg, {Processing Social Media Messages in Mass Emergency: A Survey}, ACM Computing Surveys 47~(4) (2015).
\newblock \href {http://arxiv.org/abs/arXiv:1407.7071v1} {\path{arXiv:arXiv:1407.7071v1}}, \href {https://doi.org/10.1145/2771588} {\path{doi:10.1145/2771588}}.

\bibitem{Li2012a}
C.~Li, A.~Sun, A.~Datta, {Twevent: segment-based event detection from tweets}, in: Proceedings of the 21st ACM international conference on Information and knowledge management - CIKM '12, ACM Press, New York, New York, USA, 2012, p. 155.
\newblock \href {https://doi.org/10.1145/2396761.2396785} {\path{doi:10.1145/2396761.2396785}}.

\bibitem{Mathioudakis2010}
M.~Mathioudakis, N.~Koudas, {Twittermonitor: trend detection over the twitter stream}, SIGMOD '10 (2010) 1155--1158\href {https://doi.org/10.1145/1807167.1807306} {\path{doi:10.1145/1807167.1807306}}.

\bibitem{Marcus2011}
A.~Marcus, M.~Bernstein, O.~Badar, D.~Karger, S.~Madden, R.~Miller, {Twitinfo: aggregating and visualizing microblogs for event exploration}, CHI '11  227\href {https://doi.org/10.1145/1978942.1978975} {\path{doi:10.1145/1978942.1978975}}.

\bibitem{Ritter2012}
A.~Ritter, Mausam, O.~Etzioni, S.~Clark, {Open domain event extraction from twitter}, SIGKDD'12 (2012) 1104\href {https://doi.org/10.1145/2339530.2339704} {\path{doi:10.1145/2339530.2339704}}.

\bibitem{Rogstadius2013}
J.~Rogstadius, M.~Vukovic, C.~A. Teixeira, V.~Kostakos, E.~Karapanos, J.~A. Laredo, {CrisisTracker: Crowdsourced social media curation for disaster awareness}, IBM Journal of Research and Development 57~(5) (2013) 1--4.
\newblock \href {https://doi.org/10.1147/JRD.2013.2260692} {\path{doi:10.1147/JRD.2013.2260692}}.

\bibitem{Bur2011}
K.~B{\"{u}}r, M.~Kihl, {Evaluation of selective broadcast algorithms for safety applications in vehicular ad hoc networks}, International Journal of Vehicular Technology 2011 (2011).
\newblock \href {https://doi.org/10.1155/2011/730895} {\path{doi:10.1155/2011/730895}}.

\bibitem{Goyal2012}
D.~Goyal, M.~R. Tripathy, {Routing protocols in wireless sensor networks: A survey}, ACCT'12 (2012) 474--480\href {https://doi.org/10.1109/ACCT.2012.98} {\path{doi:10.1109/ACCT.2012.98}}.

\bibitem{Kokuti2012}
A.~Kokuti, V.~Simon, {Location based data dissemination in mobile ad hoc networks}, TSP'12 (2012) 57--61\href {https://doi.org/10.1109/TSP.2012.6256252} {\path{doi:10.1109/TSP.2012.6256252}}.

\bibitem{Tonguz2006}
O.~Tonguz, N.~Wisitpongphan, J.~Parikh, F.~Bai, P.~Mudalige, V.~Sadekar, {On the broadcast storm problem in ad hoc wireless networks}, BROADNETS'06 (2006).
\newblock \href {https://doi.org/10.1109/BROADNETS.2006.4374403} {\path{doi:10.1109/BROADNETS.2006.4374403}}.

\bibitem{Heinzelman1999}
W.~Heinzelman, J.~Kulik, H.~Balakrishnan, {Adaptive protocols for information dissemination in wireless sensor networks}, MobiCom '99 (1999) 174--185\href {https://doi.org/10.1145/313451.313529} {\path{doi:10.1145/313451.313529}}.

\bibitem{Kulik2002}
J.~Kulik, W.~Heinzelman, H.~Balakrishnan, {Negotiation-based protocols for disseminating information in wireless sensor networks}, Wireless Networks 8~(2-3) (2002) 169--185.
\newblock \href {https://doi.org/10.1023/A:1013715909417} {\path{doi:10.1023/A:1013715909417}}.

\bibitem{Rehena2011}
Z.~Rehena, S.~Roy, N.~Mukherjee, {A modified SPIN for wireless sensor networks}, COMSNETS (2011) 1--4\href {https://doi.org/10.1109/COMSNETS.2011.5716469} {\path{doi:10.1109/COMSNETS.2011.5716469}}.

\bibitem{Li2014}
H.~Li, K.~Bok, K.~Chung, J.~Yoo, {An Efficient Data Dissemination Method over Wireless Ad-Hoc Networks}, Wireless Personal Communications 79~(4) (2014) 2531--2550.
\newblock \href {https://doi.org/10.1007/s11277-014-1670-x} {\path{doi:10.1007/s11277-014-1670-x}}.

\bibitem{Saleet2007}
H.~Saleet, O.~Basir, {Location-based message aggregation in vehicular ad hoc networks}, GLOBECOM (2007) 1--7\href {https://doi.org/10.1109/GLOCOMW.2007.4437821} {\path{doi:10.1109/GLOCOMW.2007.4437821}}.

\bibitem{Tsai2012}
H.~W. Tsai, {Aggregating data dissemination and discovery in vehicular Ad Hoc networks}, Telecommunication Systems 50~(4) (2012) 285--295.
\newblock \href {https://doi.org/10.1007/s11235-010-9404-1} {\path{doi:10.1007/s11235-010-9404-1}}.

\bibitem{Conti2010a}
M.~Conti, M.~Kumar, {Opportunities in opportunistic computing} (2010).

\bibitem{Meier2002}
R.~Meier, V.~Cahill, {Steam: Event-based middleware for wireless ad hoc networks}, in: Distributed Computing Systems Workshops, 2002. Proceedings. 22nd International Conference on, IEEE, 2002, pp. 639--644.

\bibitem{Koukoumidis2011}
E.~Koukoumidis, L.~Peh, M.~Martonosi, {SignalGuru: Leveraging Mobile Phones for Collaborative Traffic Signal Schedule Advisory}, in: Proceedings of the 9th international conference on Mobile systems, applications, and services, 2011, pp. 127--140.

\bibitem{Castro-Jul2017a}
F.~Castro-Jul, A.~Fernandez-Vilas, R.~P. Diaz-Redondo, {How Should My Device Behave Now? Adapting Consensus Protocols for Autonomous Context Management}, Journal of Computers 12~(3) (2017) 200--211.
\newblock \href {https://doi.org/10.17706/jcp.12.3.200-211} {\path{doi:10.17706/jcp.12.3.200-211}}.

\bibitem{Miluzzo2010}
E.~Miluzzo, C.~T. Cornelius, A.~Ramaswamy, T.~Choudhury, Z.~Liu, A.~T. Campbell, {Darwin Phones: the Evolution of Sensing and Inference on Mobile Phones}, in: Proceedings of the 8th international conference on Mobile systems, applications, and services, ACM, 2010, pp. 5--20.

\bibitem{Qin2014}
C.~Qin, X.~Bao, R.~R. Choudhury, S.~Nelakuditi, {TagSense: Leveraging Smartphones for Automatic Image Tagging}, IEEE Transactions on Mobile Computing 13~(1) (2014) 61--74.

\bibitem{Lee2012}
Y.~Lee, Y.~Ju, C.~Min, S.~Kang, I.~Hwang, J.~Song, {CoMon: Cooperative Ambience Monitoring Platform with Continuity and Benefit Awareness}, in: Proceedings of the 10th international conference on Mobile systems, applications, and services, 2012, pp. 43--56.

\bibitem{Shi2011}
C.~Shi, V.~Lakafosis, M.~H. Ammar, E.~W. Zegura, {Serendipity: Enabling Remote Computing among Intermittently Connected Mobile Devices}, in: Proceedings of the thirteenth ACM international symposium on Mobile Ad Hoc Networking and Computing, 2012, pp. 145--154.

\bibitem{Harkous2011}
H.~Harkous, J.~Makhlouta, F.~Hutayt, H.~Artail, {ACCOP: Adaptive Cost-Constrained and Delay- Optimized Data Allocation over Parallel Opportunistic Networks}, 2011, pp. 70--75.

\bibitem{OpenStreetMap2015}
OpenStreetMap, http://wiki.openstreetmap.org/ (2015).

\bibitem{Krajzewicz2012}
D.~Krajzewicz, J.~Erdmann, M.~Behrisch, L.~Bieker, {Recent development and applications of SUMO simulation of urban mobility}, International Journal On Advances in Systems and Measurements 5~(3{\&}4) (2012) 128--138.

\bibitem{ns3}
{NS - 3 network simulator}, \href{https://www.nsnam.org/}{https://www.nsnam.org/} (2015).
\newline\urlprefix\url{https://www.nsnam.org/}

\end{thebibliography}
\bibliographystyle{elsarticle-harv}

\end{document}